\documentstyle[12pt,a4]{article}
\input epsf
\global\arraycolsep=2pt

\begin{document}

\begin{titlepage}

\begin{flushright}
CERN-TH/95-112\\
hep-ph/9509432\\
October 1995
\end{flushright}

\vspace{0.5cm}

\begin{center}
\Large\bf QCD Analysis of Hadronic $\tau$ Decays Revisited
\end{center}

\vspace{1.0cm}

\begin{center}
Matthias Neubert\\
{\sl Theory Division, CERN, CH-1211 Geneva 23, Switzerland}
\end{center}

\vspace{1.2cm}

\begin{abstract}
The calculation of perturbative corrections to the spectral moments
observable in hadronic $\tau$ decays is reconsidered. The exact
order-$\alpha_s^3$ results and the resummation procedure of
Le~Diberder and Pich are compared with a partial resummation of the
perturbative series based on the analysis of so-called renormalon
chains. The perturbative analysis is complemented by a
model-independent description of power corrections. For the
contributions of dimension four and six in the OPE, it is
demonstrated how infrared renormalon ambiguities in the definition of
perturbation theory can be absorbed by a redefinition of
nonperturbative parameters. We find that previous determinations of
QCD parameters from a measurement of spectral moments in $\tau$
decays have underestimated the theoretical uncertainties. Given the
present understanding of the asymptotic behaviour of perturbation
theory, the running coupling constant can be measured at best with a
theoretical uncertainty $\delta\alpha_s(m_\tau^2)\simeq 0.05$, and
the gluon condensate with an uncertainty of order its magnitude. Two
weighted integrals of the hadronic spectral function are constructed,
which can be used to test the absence of dimension-two operators and
to measure directly the gluon condensate.
\end{abstract}

\vspace{1.0cm}

\centerline{(Submitted to Nuclear Physics B)}

\vspace{2.0cm}

\noindent
CERN-TH/95-112\\
October 1995

\end{titlepage}

\section{Introduction}

The $\tau$ lepton is the only known lepton heavy enough to decay into
hadrons. Its decays provide a unique environment to study hadronic
weak interactions at low energies. Because of its inclusive
character, the total $\tau$ hadronic width is expected to be
calculable in QCD using analyticity and the operator product
expansion (OPE) \cite{Schil}--\cite{Rtau2}.\footnote{The
applicability of the OPE in $\tau$ decays has, however, been
questioned recently \protect\cite{Shif}.}
Since nonperturbative contributions to the total width turn out to be
strongly suppressed, a measurement of the $\tau$ width is considered
an excellent way to extract the strong coupling constant $\alpha_s$
with high precision at low energies. It has also been pointed out
that hadronic $\tau$ decays can more generally be used to test QCD
and to extract some of its nonperturbative parameters
\cite{test1}--\cite{LP2}. The idea is to measure spectral moments
built from weighted integrals of the invariant hadronic mass
distribution. As the total width, such moments are calculable in QCD,
but they are more sensitive to nonperturbative contributions. An
analysis of the spectral moments can also be used to search for
non-standard effects, such as power corrections of order
$(\Lambda/m_\tau)^2$ \cite{Nari,Domi}, which are absent in the
standard OPE approach of Shifman, Vainshtein and Zakharov (SVZ)
\cite{SVZ}.

A QCD-based analysis of power-suppressed effects by measuring
spectral moments in hadronic $\tau$ decays has been pursued by two
experimental groups \cite{ALEPH,CLEO}. There is a theoretical
obstacle to any such analysis, however. The separation of
perturbative and nonperturbative effects in QCD is intrinsically
ambiguous. In order to define and extract power corrections in a
meaningful way, one has to control the perturbative contributions to
sufficiently high accuracy. But it is known that QCD perturbation
theory provides an asymptotic expansion, which is factorially
divergent at large orders. Related to this behaviour are the
so-called renormalon singularities in the Borel transform of a
perturbative series with respect to $1/\alpha_s$
\cite{tHof}--\cite{Bene}. The definition of the (resummed)
perturbative expansion by itself is ambiguous; only the sum of
perturbative and nonperturbative contributions in the OPE is well
defined \cite{Muel}. Recently, the study of renormalons has received
renewed attention. Efficient techniques have been developed to resum
so-called renormalon chains, i.e.\ terms of order
$\beta_0^{n-1}\alpha_s^n$ (where $\beta_0$ denotes the first
coefficient of the $\beta$-function), to all orders in perturbation
theory \cite{BBnew}--\cite{Turn}. These investigations are
interesting, since they provide an estimate of the importance of
higher-order terms in a series, and moreover they elucidate the
structure of nonperturbative contributions, which have to be included
in the OPE in order to cancel the ambiguities of resummed
perturbation theory. They are also useful for estimating the
uncertainty in finite-order perturbative calculations.

In the present paper, we reconsider the theoretical analysis of
hadronic $\tau$ decays in the light of these developments. This
extends some recent analyses of renormalon contributions to the
$\tau$ hadronic width \cite{part1}--\cite{Turn}. In Sect.~\ref{sec:2}
we define the spectral moments and discuss existing perturbative
predictions for them. In Sects.~\ref{sec:3} and \ref{sec:4} we
discuss the resummation of renormalon contributions for euclidean
current correlators and for the spectral moments. Section~\ref{sec:5}
is devoted to a model-independent analysis of power corrections. In
Sect.~\ref{sec:6} we show in detail how the ambiguities of
perturbation theory, which are due to the first two infrared (IR)
renormalons, can be absorbed into a redefinition of some
nonperturbative parameters. A discussion of our results and a study
of the feasibility of extracting the running coupling constant and
some nonperturbative parameters from data are presented in
Sect.~\ref{sec:7}. There we construct two weighted integrals of the
hadronic spectral function, which can be used to test the absence of
dimension-two operators in QCD and to measure the gluon condensate.
Technical details of our calculations are relegated to three
appendices.

\section{Spectral moments}
\label{sec:2}

The theoretical description of inclusive hadronic $\tau$ decays
involves two-point correlation functions of flavour-changing vector
and axial vector currents, $V_{ij}^\mu=\bar q_i\,\gamma^\mu q_j$ and
$A_{ij}^\mu=\bar q_i\,\gamma^\mu\gamma_5\,q_j$, where $i$ and $j$ are
flavour labels ($i=u$, $j=d$ or $s$). These correlators admit the
Lorentz decomposition ($\Gamma=V$ or $A$)
\begin{equation}\label{2point}
   i\int{\rm d}^4 x\,e^{i q\cdot x}\,\langle 0|\,T\,\{
   \Gamma_{ij}^\mu(x), \Gamma_{ij}^\nu(0)^\dagger\}\,|0\rangle
   = (q^\mu q^\nu - g^{\mu\nu} q^2)\,\Pi_{ij,\Gamma}^{(1)}(q^2)
   + q^\mu q^\nu\,\Pi_{ij,\Gamma}^{(0)}(q^2) \,.
\end{equation}
The superscript $(J)$ denotes the angular momentum in the rest frame
of the hadronic final state. The total hadronic $\tau$ decay rate,
normalized to the electronic one, can be written as
\begin{equation}\label{Rtaudef}
   R_\tau = {\Gamma(\tau\to\nu_\tau + \mbox{hadrons})\over
   \Gamma(\tau\to\nu_\tau\,e\,\bar\nu_e)}
   = \int\limits_0^{m_\tau^2}\!{\rm d}s\,
   {{\rm d}R_\tau(s)\over{\rm d}s} \,,
\end{equation}
where
\begin{equation}
   {{\rm d}R_\tau(s)\over{\rm d}s}
   = {24\pi S_{\rm EW}\over m_\tau^2}\,\bigg( 1 - {s\over m_\tau^2}
   \bigg)^2 \bigg[ \bigg( 1 + {2 s\over m_\tau^2} \bigg)\,
   \mbox{Im}\,\Pi^{(0+1)}(s-i\epsilon) - {2 s\over m_\tau^2}\,
   \mbox{Im}\,\Pi^{(0)}(s-i\epsilon) \bigg]
\end{equation}
is the inclusive hadronic spectrum, and
\begin{equation}\label{Picomb}
   \Pi^{(J)}(s) = {1\over 2}\,\Big( |\,V_{ud}|^2\,
   \Pi_{ud,V+A}^{(J)}(s) + |\,V_{us}|^2\,\Pi_{us,V+A}^{(J)}(s)
   \Big) \,.
\end{equation}
The factor $S_{\rm EW}\simeq 1.0194$ accounts for electroweak
radiative corrections \cite{Sirl}.

Le Diberder and Pich have pointed out the usefulness of considering
weighted integrals of the spectrum ${\rm d}R_\tau/{\rm d}s$
\cite{LP2}. Such integrals are linear combinations of the spectral
moments
\begin{equation}
   {\cal M}_k^{(J)} = {4\pi(k+1)\over m_\tau^{2k+2}}\,
   \int\limits_0^{m_\tau^2}\!{\rm d}s\,s^k\,\mbox{Im}\,
   \Pi^{(J)}(s-i\epsilon) \,;\quad k\ge 0 \,,
\end{equation}
which contain the dynamical information that can be extracted from
hadronic $\tau$ decays.\footnote{Here we shall not pursue the idea to
disentangle vector from axial vector, or strange from non-strange
modes, although in principle this may be possible.}
The weighted integrals
\begin{equation}
   R_k = {(k+1)(k+3)(k+4)\over 36 S_{\rm EW}}\,
   \int\limits_0^{m_\tau^2}\!{\rm d}s\,\bigg( {s\over m_\tau^2}
   \bigg)^k\,{{\rm d}R_\tau(s)\over{\rm d}s} \,,
\end{equation}
which are normalized such that their perturbative expansion is
$R_k=1+\alpha_s/\pi+\dots$, are given by
\begin{eqnarray}\label{Rkmess}
   R_k &=& {(k+3)(k+4)\over 6}\,{\cal M}_k^{(0+1)}
    - {(k+1)(k+4)\over 2}\,{\cal M}_{k+2}^{(0+1)}
    + {(k+1)(k+3)\over 3}\,{\cal M}_{k+3}^{(0+1)} \nonumber\\
   &&\mbox{}- {(k+1)(k+3)(k+4)\over 3(k+2)}\,{\cal M}_{k+1}^{(0)}
    + {2(k+1)(k+4)\over 3}\,{\cal M}_{k+2}^{(0)}
    - {(k+1)(k+3)\over 3}\,{\cal M}_{k+3}^{(0)} \,.
    \nonumber\\
\end{eqnarray}
In particular, we note that
\begin{equation}\label{Rtau}
   R_0 = {R_\tau\over 3 S_{\rm EW}} = 2 {\cal M}_0^{(0+1)}
   - 2 {\cal M}_2^{(0+1)} + {\cal M}_3^{(0+1)}
   - 2 {\cal M}_1^{(0)} + {8\over 3}\,{\cal M}_2^{(0)}
   - {\cal M}_3^{(0)} \,.
\end{equation}

Using the analyticity properties of the correlators in
(\ref{2point}), the spectral moments can be written as contour
integrals in the complex $s$-plane. One finds
\begin{equation}\label{Mcirc}
   {\cal M}_k^{(J)} = {1\over 2\pi i}\,\oint\limits_{|s|=m_\tau^2}
   {{\rm d}s\over s}\,\bigg[ 1 - \bigg( {s\over m_\tau^2}
   \bigg)^{k+1} \bigg]\,D^{(J)}(-s) \,,
\end{equation}
where
\begin{equation}\label{Dijdef}
   D^{(J)}(-s) = 4\pi^2 s\,{{\rm d}\over{\rm d}s}\,\Pi^{(J)}(s)
\end{equation}
are the logarithmic derivatives of the correlators, which are
ultraviolet (UV) finite. These functions are analytic in the complex
$s$-plane, with discontinuities on the positive real axis. The
contour integrals have two attractive features. First, to perform the
integration along a circle with radius $|s|=m_\tau^2$ requires
knowledge of the correlation functions only for large (complex)
momenta. Second, the integrand in (\ref{Mcirc}) vanishes for
$s=m_\tau^2$, where the contour touches the physical cut. Therefore,
unlike the spectral functions themselves, their moments are expected
to admit a theoretical description in QCD, which can be organized as
an expansion in powers of $(\Lambda/m_\tau)^{2n}$ \cite{Rtau1,LP2}.

One of the purposes of this paper is to reconsider the calculation of
the perturbative contribution in this expansion, i.e.\ the term with
$n=0$. This contribution can be calculated by setting the current
quark masses of the light quarks $u$, $d$ and $s$ to zero. In this
limit the flavour-changing currents are conserved, so that the
correlators with $J=0$ in (\ref{2point}) vanish. Moreover, chiral
invariance implies that the spectral functions in the vector and
axial vector channels are the same. It follows that\footnote{We set
$|\,V_{ud}|^2 + |\,V_{us}|^2 = 1$, which is an excellent
approximation, since experimentally $|\,V_{ub}|^2<1.6\times
10^{-5}$.}
\begin{equation}
   D_{\rm pert}^{(0+1)}(-s) = D(-s) \,,\qquad
   D_{\rm pert}^{(0)}(-s) = 0 \,.
\end{equation}
The perturbative expansion of the function $D(-s)$,
\begin{equation}\label{Dseries}
   D(-s) = 1 + \sum_{n=1}^\infty\,K_n\,\bigg( {\alpha_s(-s)\over\pi}
   \bigg)^n \,,
\end{equation}
is known to order $\alpha_s^3$. For $n\ge 3$, the coefficients $K_n$
depend on the renormalization scheme. In the $\overline{\rm MS}$
scheme, one finds (with $n_f=3$ quark flavours)
\cite{Ree1}--\cite{Tals3b}
\begin{eqnarray}\label{K1K2K3}
   K_1 &=& 1 \,, \nonumber\\
   K_2 &=& {299\over 24} - 9\zeta(3) \simeq 1.63982 \,, \nonumber\\
   K_3^{\overline{\rm MS}} &=& {58057\over 288} - {779\over 4}\,
    \zeta(3) + {75\over 2}\,\zeta(5) \simeq 6.37101 \,,
\end{eqnarray}
with $\zeta(3)\simeq 1.20206$ and $\zeta(5)\simeq 1.03693$. The next
coefficient, $K_4$, has been estimated using the principle of minimal
sensitivity \cite{PMS} and the effective charge approach \cite{ECH},
with the result that \cite{Kata}
\begin{equation}\label{K4}
    K_{4,{\rm est}}^{\overline{\rm MS}}\simeq 27.5 \,.
\end{equation}

The perturbative contributions to the moments in (\ref{Mcirc}) are
\begin{equation}
   {\cal M}_{k,{\rm pert}}^{(0+1)} = M_k \,,\qquad
   {\cal M}_{k,{\rm pert}}^{(0)} = 0 \,,
\end{equation}
with
\begin{equation}\label{circle}
   M_k = {1\over 2\pi i}\,\oint\limits_{|x|=1} {{\rm d}x\over x}\,
   (1 - x^{k+1})\,D(-x m_\tau^2) = 1 + \sum_{n=1}^\infty\,d_n\,
   \bigg( {\alpha_s(m_\tau^2)\over\pi} \bigg)^n \,.
\end{equation}
The coefficients $d_n$ can be obtained using the
renormalization-group equation (RGE)
\begin{equation}\label{RGE}
   \mu^2\,{{\rm d}\alpha_s(\mu^2)\over{\rm d}\mu^2}
   = -\alpha_s(\mu^2)\,\beta[\alpha_s(\mu^2)] = -\alpha_s(\mu^2)\,
   \sum_{n=0}^\infty\,\beta_n\,\bigg( {\alpha_s(\mu^2)\over 4\pi}
   \bigg)^{n+1}
\end{equation}
for the running coupling constant. For $n_f=3$, the first three
coefficients of the $\beta$-function are $\beta_0=9$, $\beta_1=64$
and $\beta_2^{\overline{\rm MS}}=3863/6$ \cite{beta2}. The result is
\cite{LP2}
\begin{eqnarray}\label{dcoef}
   d_1 &=& 1 \,, \nonumber\\
   d_2 &=& K_2 + {\beta_0\over 4(k+1)}
    \simeq 1.63982 + {2.25\over k+1} \,, \nonumber\\
   d_3^{\overline{\rm MS}} &=& K_3^{\overline{\rm MS}}
    + {1\over 2(k+1)}\,\bigg( \beta_0 K_2 + {\beta_1\over 8} \bigg)
    + {\beta_0^2\over 8}\,\bigg( {1\over(k+1)^2} - {\pi^2\over 6}
    \bigg) \nonumber\\
   &\simeq& -10.2839 + {11.3792\over k+1} + {10.125\over(k+1)^2} \,,
    \nonumber\\
   d_4^{\overline{\rm MS}} &=& K_4^{\overline{\rm MS}}
    + \bigg( {3\over 8(k+1)^2} - {\pi^2\over 16} \bigg)
    \bigg( \beta_0^2 K_2 + {5\over 24}\,\beta_0\beta_1 \bigg)
    \nonumber\\
   &&\mbox{}+ {3\over 4(k+1)}\,\bigg( \beta_0 K_3^{\overline{\rm MS}}
    + {\beta_1\over 6}\,K_2 + {\beta_2^{\overline{\rm MS}}\over 48}
    - {\beta_0^3\,\pi^2\over 48} \bigg) + {3\beta_0^3\over 32(k+1)^3}
    \nonumber\\
   &\simeq& K_4^{\overline{\rm MS}}
    - 155.955 - {46.238\over k+1} + {94.810\over(k+1)^2}
    + {68.344\over(k+1)^3} \,.
\end{eqnarray}
As mentioned above, the exact result for the coefficient
$K_4^{\overline{\rm MS}}$ is unknown. Nevertheless, it follows that
ratios and differences of moments are known exactly to order
$\alpha_s^4$, since this unknown coefficient cancels out. In
Table~\ref{tab:1}, we show the order-$\alpha_s^3$ predictions for the
first six moments as a function of the value of $\alpha_s(m_\tau^2)$
in the $\overline{\rm MS}$ scheme.

\begin{table}[t]
\centerline{\parbox{15cm}{\caption{\label{tab:1}\protect\small\sc
Perturbative contributions to the moments at order $\alpha_s^3$.}}}
\vspace{0.5cm}
\centerline{\begin{tabular}{c|cccccc}
\hline\hline
\rule[-0.2cm]{0cm}{0.7cm} $\alpha_s(m_\tau^2)$ & $M_0$ & $M_1$ &
 $M_2$ & $M_3$ & $M_4$ & $M_5$ \\
\hline
0.26 & 1.116 & 1.101 & 1.096 & 1.094 & 1.093 & 1.092 \\
0.28 & 1.128 & 1.110 & 1.104 & 1.102 & 1.100 & 1.099 \\
0.30 & 1.141 & 1.119 & 1.113 & 1.110 & 1.108 & 1.107 \\
0.32 & 1.154 & 1.128 & 1.121 & 1.118 & 1.116 & 1.114 \\
0.34 & 1.168 & 1.138 & 1.129 & 1.125 & 1.123 & 1.122 \\
0.36 & 1.183 & 1.148 & 1.138 & 1.133 & 1.131 & 1.129 \\
0.38 & 1.198 & 1.158 & 1.146 & 1.141 & 1.138 & 1.136 \\
\hline\hline
\end{tabular}}
\vspace{0.5cm}
\end{table}

Le Diberder and Pich have argued that one should improve the
perturbative prediction for the spectral moments by performing a
partial resummation of higher-order terms \cite{LP1,LP2}. Their
observation was that the evolution of the running coupling constant
along the integration contour in (\ref{circle}) generates
higher-order corrections that are expected to be larger than the
``genuine'' higher-order corrections to the correlator $D(-s)$. These
corrections can be resummed by writing
\begin{eqnarray}\label{Pich}
   M_k^{\rm LP} &=& 1 + \sum_{n=1}^3\,K_n\,I_n^{k+1}(m_\tau) \,,
    \nonumber\\
   I_n^{k+1}(m_\tau) &=& {1\over 2\pi i}\,\oint\limits_{|x|=1}
    {{\rm d}x\over x}\,(1 - x^{k+1})\,\bigg(
    {\alpha_s(-x m_\tau^2)\over\pi} \bigg)^n \,.
\end{eqnarray}
The contour integrals over the running coupling constant can be
performed numerically (see Appendix~\ref{app:int}). The effect of
this resummation is significant. In Table~\ref{tab:2}, we show the
results obtained using the two-loop RGE for the running coupling
constant. The differences with respect to fixed-order perturbation
theory are of order a few per cent. We shall comment on the
usefulness of this resummation procedure at the end of
Sect.~\ref{sec:4}.

\begin{table}[t]
\centerline{\parbox{15cm}{\caption{\label{tab:2}\protect\small\sc
Perturbative contributions to the moments obtained from the
resummation procedure of Le~Diberder and Pich (LP)
\protect\cite{LP2}.}}}
\vspace{0.5cm}
\centerline{\begin{tabular}{c|cccccc}
\hline\hline
\rule[-0.2cm]{0cm}{0.7cm} $\alpha_s(m_\tau^2)$ & $M_0^{\rm LP}$ &
 $M_1^{\rm LP}$ & $M_2^{\rm LP}$ & $M_3^{\rm LP}$ & $M_4^{\rm LP}$ &
 $M_5^{\rm LP}$ \\
\hline
0.26 & 1.110 & 1.090 & 1.087 & 1.086 & 1.085 & 1.084 \\
0.28 & 1.120 & 1.095 & 1.093 & 1.091 & 1.090 & 1.089 \\
0.30 & 1.130 & 1.100 & 1.098 & 1.096 & 1.095 & 1.094 \\
0.32 & 1.140 & 1.105 & 1.103 & 1.101 & 1.100 & 1.099 \\
0.34 & 1.150 & 1.109 & 1.107 & 1.105 & 1.104 & 1.103 \\
0.36 & 1.160 & 1.112 & 1.112 & 1.109 & 1.108 & 1.107 \\
0.38 & 1.169 & 1.114 & 1.116 & 1.113 & 1.111 & 1.111 \\
\hline\hline
\end{tabular}}
\vspace{0.5cm}
\end{table}

\begin{figure}[htb]
   \vspace{0.5cm}
   \epsfysize=3cm
   \centerline{\epsffile{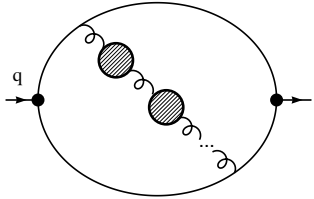}}
   \centerline{\parbox{13cm}{\caption{\label{fig:renchain}
\protect\small\sc
Renormalon-chain contribution to the correlator $D(-q^2)$. The shaded
bubble represents a self-energy insertion on the gluon propagator. We
note that bubble summation is not a gauge-invariant procedure in a
non-abelian theory; the figure is meant as an illustration only.
}}}
\end{figure}

In this paper we investigate another type of resummation, which deals
with the so-called renormalon-chain contributions
\cite{tHof}--\cite{Turn}. These are terms of order
$\beta_0^{n-1}\alpha_s^n$ in the perturbative series for $D(-s)$ and
for the moments $M_k$, where $\beta_0=11-\frac{2}{3}\,n_f$ is the
first coefficient of the $\beta$-function. A graphical representation
of a renormalon chain is shown in Fig.~\ref{fig:renchain}. The
motivation to resum these terms to all orders is the following: The
coefficients in the perturbative series (\ref{Dseries}) and
(\ref{circle}), which are polynomials in the number of quark flavours
$n_f$, can be reorganized as polynomials in $\beta_0$, i.e.
\begin{eqnarray}\label{kappan}
   K_n &=& \kappa_n\,\beta_0^{n-1} + \kappa_n^{(n-2)}\,
    \beta_0^{n-2} + \dots + \kappa_n^{(1)}\,\beta_0
    + \kappa_n^{(0)} \,, \nonumber\\
   d_n &=& \delta_n\,\beta_0^{n-1} + \delta_n^{(n-2)}\,
    \beta_0^{n-2} + \dots + \delta_n^{(1)}\,\beta_0
    + \delta_n^{(0)} \,.
\end{eqnarray}
Since in QCD the value of $\beta_0$ is large, it is conceivable that
a reasonable approximation to these coefficients is provided by the
first term, i.e.\ the one with the highest power of $\beta_0$. Some
formal arguments supporting this assertion can be found in
Ref.~\cite{Turn}. Below, we shall refer to this approximation as the
``large-$\beta_0$ limit''. In this limit, one finds
\begin{eqnarray}
   \kappa_1 &=& 1 \,, \nonumber\\
   \kappa_2\,\beta_0 &=& {297\over 24} - 9 \zeta(3)\simeq 1.55649 \,,
    \nonumber\\
   \kappa_3\,\beta_0^2 &=& {48924\over 288} - {513\over 4}\,
    \zeta(3)\simeq 15.7112 \,, \nonumber\\
   \kappa_4\,\beta_0^3 &=& {165537\over 64} - {49329\over 32}\,
    \zeta(3) - {10935\over 16}\,\zeta(5)\simeq 24.8320 \,,
\end{eqnarray}
which can be compared with the exact values of $K_1$, $K_2$ and
$K_3^{\overline{\rm MS}}$ given in (\ref{K1K2K3}), as well as with
the estimated value of $K_4^{\overline{\rm MS}}$ given in (\ref{K4}).
With the exception of $K_3^{\overline{\rm MS}}$, the large-$\beta_0$
limit seems to work quite well. We also quote the corresponding
results for the coefficients in the expansion of the moments. They
are
\begin{eqnarray}\label{largeb0}
   \delta_1 &=& 1 \,, \nonumber\\
   \delta_2\,\beta_0 &=& \bigg( {11\over 8} - \zeta(3)
    + {1\over 4(k+1)} \bigg)\,\beta_0 \simeq 1.55649
    + {2.25\over k+1} \,, \nonumber\\
   \delta_3\,\beta_0^2 &=& \Bigg( {151\over 72} - {19\over 12}\,
    \zeta(3) - {\pi^2\over 48} + \bigg( {11\over 16} - {1\over 2}\,
    \zeta(3) \bigg)\,{1\over k+1} + {1\over 8(k+1)^2}
    \Bigg)\,\beta_0^2 \nonumber\\
   &\simeq& -0.9442 + {7.0041\over k+1} + {10.125\over(k+1)^2} \,,
\end{eqnarray}
which can be compared with the exact values of $d_1$, $d_2$ and
$d_3^{\overline{\rm MS}}$ given in (\ref{dcoef}).

In Table~\ref{tab:3}, we show the results obtained in the
large-$\beta_0$ limit for the perturbative contributions to the
moments at order $\alpha_s^3$. For the difference between the
``large-$\beta_0$ results'' and the exact ones, we find
\begin{equation}\label{correct}
   M_k - M_k^{{\rm large}-\beta_0} = {1\over 12}\,\bigg(
   {\alpha_s\over\pi} \bigg)^2 + \bigg( {9133\over 288}
   - {133\over 2}\,\zeta(3) + {75\over 2}\,\zeta(5)
   + {35\over 8(k+1)} \bigg) \bigg( {\alpha_s\over\pi} \bigg)^3
   + O(\alpha_s^4) \,,
\end{equation}
where $\alpha_s=\alpha_s(m_\tau^2)$ in the $\overline{\rm MS}$
scheme. A comparison of Tables~\ref{tab:1} and \ref{tab:3} shows that
these differences are rather small, supporting to some extent the
usefulness of the large-$\beta_0$ limit.

\begin{table}[t]
\centerline{\parbox{15cm}{\caption{\label{tab:3}\protect\small\sc
Large-$\beta_0$ approximation to the perturbative contributions to
the moments at order $\alpha_s^3$.}}}
\vspace{0.5cm}
\centerline{\begin{tabular}{c|cccccc}
\hline\hline
\rule[-0.2cm]{0cm}{0.7cm} $\alpha_s(m_\tau^2)$ &
 $M_0^{{\rm large}-\beta_0}$ & $M_1^{{\rm large}-\beta_0}$ &
 $M_2^{{\rm large}-\beta_0}$ & $M_3^{{\rm large}-\beta_0}$ &
 $M_4^{{\rm large}-\beta_0}$ & $M_5^{{\rm large}-\beta_0}$ \\
\hline
0.26 & 1.118 & 1.104 & 1.100 & 1.098 & 1.097 & 1.096 \\
0.28 & 1.131 & 1.114 & 1.109 & 1.107 & 1.106 & 1.105 \\
0.30 & 1.144 & 1.124 & 1.119 & 1.116 & 1.115 & 1.114 \\
0.32 & 1.158 & 1.135 & 1.128 & 1.125 & 1.124 & 1.122 \\
0.34 & 1.173 & 1.146 & 1.138 & 1.135 & 1.133 & 1.131 \\
0.36 & 1.189 & 1.157 & 1.149 & 1.145 & 1.142 & 1.141 \\
0.38 & 1.205 & 1.169 & 1.159 & 1.155 & 1.152 & 1.150 \\
\hline\hline
\end{tabular}}
\vspace{0.5cm}
\end{table}

In Refs.~\cite{BBnew}--\cite{Turn}, efficient techniques have been
developed to resum the terms of order $\beta_0^{n-1}\alpha_s^n$
exactly to all orders in perturbation theory. Note that such a
resummation includes the main part of the resummation of Le~Diberder
and Pich. In addition, a partial resummation of the higher-order
coefficients $K_n$ in the series (\ref{Dseries}) is achieved. This is
important, since these coefficients are known to diverge factorially
for large $n$, reflecting the asymptotic behaviour of perturbative
expansions in QCD. The perturbative series for $D(-s)$ and for the
moments $M_k$ are not Borel summable, implying that the definition of
these quantities in perturbation theory has an intrinsic ambiguity,
which has to be cured by adding nonperturbative contributions
\cite{Muel}. The resummation of renormalon-chain contributions allows
one to quantify (or at least estimate) this ambiguity. Moreover,
unlike any truncated perturbative series, the resummed expressions
for $D(-s)$ and for the moments $M_k$ are independent of the
renormalization scheme. Nevertheless, one has to keep in mind that
this method provides only a partial resummation of a perturbative
series, and it is not obvious that it gives a better approximation to
the full series than fixed-order perturbation theory. Therefore, to
be conservative we shall take the differences between our procedure
and others as an indication of the perturbative uncertainty.

\section{Renormalon resummation for the $D$ function}
\label{sec:3}

We start by considering the correlator $D(-s)$ in the euclidean
region, where $-s=Q^2>0$. The terms of order $\beta_0^{n-1}
\alpha_s^n$ in the perturbative series (\ref{Dseries}) can be
resummed by evaluating the Borel integral
\begin{eqnarray}\label{Laplace}
   D_{\rm Borel}(Q^2) &=& 1 + {4\over\beta_0}\,
    \int\limits_0^\infty\!{\rm d}u\,\widehat S_D(u)\,\exp\bigg(
    - {4\pi u\over\beta_0\alpha_s(Q^2)} \bigg) \nonumber\\
   &=& 1 + {4\over\beta_0}\,\int\limits_0^\infty\!{\rm d}u\,
    \widehat S_D(u)\,\bigg( {\Lambda^2\over Q^2} \bigg)^u \,,
\end{eqnarray}
with the Borel transform \cite{Broa,Bene}
\begin{equation}\label{SDu}
   \widehat S_D(u) = \sum_{n=1}^\infty\,{(4 u)^{n-1}\over\Gamma(n)}\,
   \kappa_n = {32\,e^{-C u}\over 3(2-u)}\,\sum_{k=2}^\infty\,
   {(-1)^k\,k\over\big[ k^2-(1-u)^2\big]^2} \,.
\end{equation}
$\Lambda$ is the scale parameter in the one-loop expression for the
running coupling constant, and $C$ is a scheme-dependent constant,
with $C=-5/3$ in the $\overline{\rm MS}$ scheme ($C=0$ in the V
scheme, see below). The coefficients $\kappa_n$ have been defined in
(\ref{kappan}). The function $\widehat S_D(u)$ contains pole
singularities on the real $u$ axis. The singularities on the negative
axis are related to the UV behaviour of Feynman diagrams and are
called UV renormalons. The singularities on the positive axis, which
arise from the integration over low virtual momenta in Feynman
diagrams, are called IR renormalons \cite{tHof}. In the present case
there are UV renormalon poles located at $u=-1,-2,\dots$ and IR
renormalon poles at $u=2,3,\dots$. In general, a pole at $u=u_i$ is
associated with a factorial growth of the expansion coefficients
$\kappa_n$ of the form $\Gamma(n)\,(4 u_i)^{-n}$ (there is an extra
factor $n$ for a double pole), so that the renormalon singularity
closest to the origin determines the asymptotic behaviour of the
expansion coefficients. In the case of UV renormalons, these
contributions have alternating sign and can be resummed by means of
the Borel integral (\ref{Laplace}). However, for IR renormalons all
terms have the same sign and the series is divergent. In fact, IR
renormalon singularities fall on the integration contour in
(\ref{Laplace}), making the value of the Borel integral ambiguous. A
measure of the ambiguity is provided by the residue uf the nearest IR
renormalon pole. In the present case this is located at $u=2$ and
leads to an ambiguity of order $\Lambda^4/Q^4$.

The appearance of IR renormalons indicates that perturbation
theory is incomplete; it must be supplemented by nonperturbative
corrections. Only the sum of all perturbative and nonperturbative
contributions is unambiguous. For euclidean correlation functions of
currents, the OPE provides a consistent framework for a systematic
incorporation of nonperturbative effects \cite{Wils}. In the case of
the function $D(Q^2)$, such effects appear first at order $1/Q^4$ and
are parametrized by the gluon condensate \cite{SVZ}, which has an
ambiguity that compensates the ambiguity in the resummed perturbative
series \cite{reno2}--\cite{Muel}. This will be discussed in more
detail in Sect.~\ref{sec:6}.

\begin{figure}[htb]
   \vspace{0.5cm}
   \epsfysize=6cm
   \centerline{\epsffile{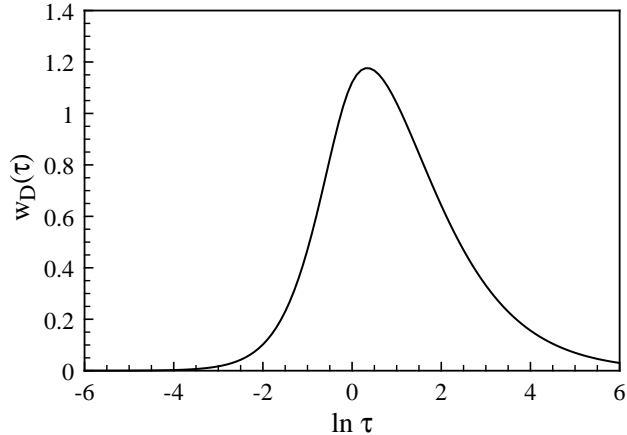}}
   \centerline{\parbox{13cm}{\caption{\label{fig:1}\protect\small\sc
Distribution function $w_D(\tau)$ versus $\ln\tau$.}}}
\end{figure}

An efficient technique for evaluating the Borel integral
(\ref{Laplace}) has been developed in Ref.~\cite{part1}, where it was
shown that\footnote{In Ref.~\protect\cite{part1}, the function
$w_D(\tau)$ was denoted $\tau\,\widehat w_D(\tau)$.}
\begin{equation}\label{Dsum}
   D_{\rm Borel}(Q^2) = 1 + {1\over\beta_0}\,\int\limits_0^\infty
   {{\rm d}\tau\over\tau}\,w_D(\tau)\,a(\tau Q^2) \,,
\end{equation}
where
\begin{equation}\label{adef}
   a(\tau Q^2) = {1\over\ln\tau + \ln(Q^2/\Lambda_{\rm V}^2)}
   = {\beta_0\over 4\pi}\,\alpha_s^{({\rm V})}(\tau Q^2) \,.
\end{equation}
Here $\alpha_s^{({\rm V})}(\tau Q^2)$ is the one-loop running
coupling constant in the so-called V scheme, in which the coupling
constant is defined in terms of the heavy-quark potential \cite{BLM}.
The relation between the V scheme and the $\overline{\rm MS}$ scheme
is such that $\alpha_s^{({\rm V})}(\mu^2)=\alpha_s(e^C \mu^2)$ with
$C=-5/3$. Thus, the parameter $\Lambda_{\rm V}$ can be obtained from
the value of $\alpha_s(m_\tau^2)$ in the $\overline{\rm MS}$ scheme
by
\begin{equation}\label{LamV}
   \Lambda_{\rm V} = m_\tau\,\exp\bigg( {5\over 6}
   - {2\pi\over 9\alpha_s(m_\tau^2)} \bigg)
   = e^{5/6}\,\Lambda_{\overline{\rm MS}} \,.
\end{equation}
The function $w_D(\tau)$ in (\ref{Dsum}) describes the distribution
of the virtuality of the gluon in the diagrams associated with the
order-$\alpha_s$ correction to $D(Q^2)$. It is given by \cite{part1}
\begin{eqnarray}\label{wDfun}
   w_D(\tau) &=& {32\over 3}\,\bigg\{
    \bigg( {7\over 4} - \ln\tau \bigg)\,\tau^2
    + \tau\,(1+\tau)\,\Big[ L_2(-\tau) + \ln\tau\,\ln(1+\tau) \Big]
    \bigg\} \,;\quad \tau<1 \,, \nonumber\\
   && \nonumber\\
   w_D(\tau) &=& {32\over 3}\,\bigg\{ {3\over 4}
    + {1\over 2}\,\ln\tau + ( 1 + \ln\tau )\,\tau
    \nonumber\\
   &&\qquad \mbox{}+ \tau\,(1+\tau)\,\Big[ L_2(-\tau^{-1})
    - \ln\tau\,\ln(1+\tau^{-1}) \Big] \bigg\} \,;\quad \tau>1 \,.
\end{eqnarray}
$L_2(x)=-\int_0^x {{\rm d}t\over t}\ln(1-t)$ is the dilogarithm. In
Fig.~\ref{fig:1}, we show $w_D(\tau)$ as a function of $\ln\tau$,
which is the natural integration variable in (\ref{Dsum}). The
expansion coefficients $\kappa_n$ are related to the moments of this
distribution by
\begin{equation}
   \kappa_n = \int\limits_0^\infty {{\rm d}\tau\over\tau}\,
   w_D(\tau)\,(-\ln\tau-C)^{n-1} \,.
\end{equation}

The value of the integral in (\ref{Dsum}) is independent of the
renormalization scheme; changing the scheme simply amounts to
rescaling the integration variable $\tau$.\footnote{This is true for
regular (MS-like) schemes only.}
However, as written above the integral is ambiguous because the
integration contour runs over the Landau pole in the running coupling
constant $a(\tau Q^2)$, which is located at $\tau=\tau_L=\Lambda_{\rm
V}^2/Q^2$. This is nothing than the IR renormalon ambiguity mentioned
above. Different regularization prescriptions for the Landau pole
lead to different results. A measure of the ambiguity is provided by
the residue of the pole, i.e.\
\begin{equation}\label{Dren}
   \Delta D_{\rm ren}(Q^2) = {1\over\beta_0}\,w_D(\tau_L)
   = {8\over\beta_0}\,{\Lambda_{\rm V}^4\over Q^4}
   - {16\over 3\beta_0}\,\bigg( \ln{Q^2\over\Lambda_{\rm V}^2}
   + {3\over 2} \bigg)\,{\Lambda_{\rm V}^6\over Q^6} + O(Q^{-8}) \,.
\end{equation}
The quantity $\Delta D_{\rm ren}(Q^2)$ is equal to the sum of the
residues of the IR renormalon poles in the integrand of the Borel
integral in (\ref{Laplace}) \cite{part1}. Below we shall use the
principle value prescription to regulate the Landau pole in
(\ref{Dsum}), which is equivalent to calculating the principle value
of the Borel integral.

To illustrate the importance of the all-order resummation, we quote
numerical results for the case $Q^2=m_\tau^2$ and
$\alpha_s(m_\tau^2)=0.32$ in the $\overline{\rm MS}$ scheme. For
these parameters, the order-$\alpha_s^3$ result in the
large-$\beta_0$ limit is $D(m_\tau^2)=1.135$ (the exact result is
1.126). The resummation of all terms of order
$\beta_0^{n-1}\alpha_s^n$ leads to a value that is significantly
larger, $D_{\rm Borel}(m_\tau^2)=1.151\pm 0.003$. The central value
corresponds to the principle value of the Borel integral, whereas the
error is given by the renormalon ambiguity $\Delta D_{\rm ren}$ in
(\ref{Dren}). It is instructive to investigate the behaviour of the
perturbative series in the large-$\beta_0$ limit. The expansion
coefficients $\kappa_n$ can be obtained by expanding the Borel
transform $\widehat S_D(u)$ in (\ref{SDu}) in powers of $u$. For the
$\overline{\rm MS}$ and the V scheme, the results are given in
Table~\ref{tab:Kn}. In the V scheme, the coefficients have
alternating sign, and the nearest UV renormalon at $u=-1$ is dominant
already at low orders \cite{Turn}. In the $\overline{\rm MS}$ scheme,
UV renormalons are not yet dominant at low orders due to an extra
factor $e^{5 u/3}$ in the Borel transform, which suppresses the
region of negative $u$. The first few coefficients have the same
sign, and their growth is determined by the interplay of the UV
renormalon at $u=-1$ and the IR renormalon at $u=2$.

\begin{table}[htb]
\centerline{\parbox{15cm}{\caption{\label{tab:Kn}\protect\small\sc
Perturbative coefficients for the $D$ function in the large-$\beta_0$
limit.}}}
\vspace{0.5cm}
\centerline{\begin{tabular}{c|ccccc}
\hline\hline
\rule[-0.2cm]{0cm}{0.7cm} $n$ & 1 & 2 & 3 & 4 & 5 \\
\hline
\rule[-0.1cm]{0cm}{0.6cm}
 $\kappa_n^{\overline{\rm MS}}\,\beta_0^{n-1}$ & 1 &
 $\phantom{-}1.55649$ & 15.7112 & $\phantom{-}24.8320$ & 787.827 \\
\rule[-0.1cm]{0cm}{0.6cm}
 $\kappa_n^{\rm V}\,\beta_0^{n-1}$ & 1 & $-2.19351$ & 18.1000 &
 $-138.989$ & 1610.41 \\
\hline\hline
\end{tabular}}
\vspace{0.5cm}
\end{table}

In Table~\ref{tab:series}, we show the partial sums $D^{(N)} = 1 +
\sum_{n=1}^N\,\kappa_n\,\beta_0^{n-1}(\alpha_s/\pi)^n$ in the two
schemes. Note that the asymptotic value, which is given by the
integral in (\ref{Dsum}), is scheme independent. In the
$\overline{\rm MS}$ scheme, the minimal term in the series is reached
at $n=4$ and gives a contribution of 0.3\%. If one truncates the
series at this point, the difference to the resummed result is 1.4\%,
which is about four times the renormalon ambiguity. In the V scheme,
the minimal term is reached already at $n=2$, and its contribution is
6.0\%. If the series is truncated at this point, the difference to
the resummed result is 4.6\%. Since in the V scheme the leading UV
renormalon is dominant, the error due to the truncation of the series
is necessarily of order $(\Lambda_{\rm V}/m_\tau)^2\simeq 6.7\%$. The
fact that this error is much larger than the IR renormalon ambiguity,
and moreover has a different power dependence on $m_\tau$, has been
emphasized in Ref.~\cite{Alta}. For comparison, we note that the
leading nonperturbative contribution to the $D$ function, which is
proportional to the gluon condensate \cite{SVZ}, is given by (see
Appendix~\ref{app:QCD})
\begin{equation}
   D_{\rm power}(m_\tau^2) = {2\pi^2\over 3 m_\tau^4}\,
   \bigg\langle {\alpha_s\over\pi}\,G^2\bigg\rangle
   \simeq (1.2\pm 0.6)\% \,.
\end{equation}
Clearly, to give meaning to the value of nonperturbative parameters
such as the vacuum condensates one has to control the higher-order
behaviour of perturbation theory; in particular, it is necessary to
resum the contribution of the first UV renormalon. If one believes
that the large-$\beta_0$ limit provides a reasonable description of
the nature of this singularity, the resummation is achieved by
performing the Borel integral, and then the residual ambiguity is due
to the nearest IR renormalon and thus of order $(\Lambda_{\rm
V}/m_\tau)^4$. If, however, one wants to be more conservative and
truncate the series at the last term that is exactly known, one is
left with an uncertainty that is larger, namely of order the last
term itself.

\begin{table}[htb]
\centerline{\parbox{15cm}{\caption{\label{tab:series}
\protect\small\sc
Partial sums of the perturbative series for the $D$ function in the
large-$\beta_0$ limit. In the second column we show for comparison
the known exact results in the $\overline{\rm MS}$ scheme.}}}
\vspace{0.5cm}
\centerline{\begin{tabular}{c|ccc}
\hline\hline
\rule[-0.2cm]{0cm}{0.7cm} $N$ &
 $D_{\overline{\rm MS}}^{(N)}(m_\tau^2)$ & exact result
($\overline{\rm MS}$) & $D_{\rm V}^{(N)}(m_\tau^2)$ \\
\hline
1 & 1.1019 & 1.1019 & 1.1648 \\
2 & 1.1180 & 1.1189 & 1.1052 \\
3 & 1.1346 & 1.1256 & 1.1863 \\
4 & 1.1373 &        & 1.0837 \\
5 & 1.1459 &        & 1.2795 \\
$\infty$ & $1.151\pm 0.003$ & & $1.151\pm 0.003$ \\
\hline\hline
\end{tabular}}
\vspace{0.5cm}
\end{table}

\section{Renormalon resummation for the moments}
\label{sec:4}

\begin{figure}[htb]
   \vspace{0.5cm}
   \epsfysize=6cm
   \centerline{\epsffile{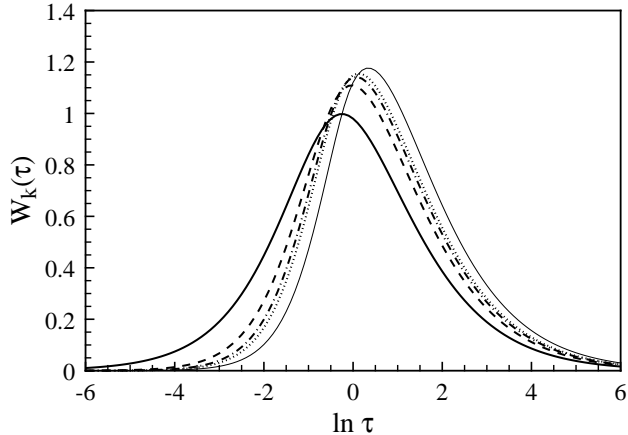}}
   \centerline{\parbox{13cm}{\caption{\label{fig:3}\protect\small\sc
Functions $W_k(\tau)$ versus $\ln\tau$, for $k=0$ (solid line), $k=1$
(dashed line), $k=2$ (dash-dotted line) and $k=3$ (dotted line). The
asymptotic result for $k\to\infty$ is shown as the thin solid
line.}}}
\end{figure}

Let us now turn to the resummation of renormalon-chain contributions
for the spectral moments. In the Minkowski region, a simple
representation of the Borel integral of the form (\ref{Dsum}) does
not exist \cite{part1}. Beneke et al.\ have shown that the principle
value of the Borel integral can be obtained from \cite{BBnew,BBB}
\begin{equation}\label{MkBorel}
   M_k^{\rm Borel} = 1 + {1\over\pi\beta_0}\,\int\limits_0^\infty
   {{\rm d}\tau\over\tau}\,W_k(\tau)\,\arctan[\pi a(\tau m_\tau^2)]
   + {1\over\beta_0}\,\mbox{Re}\,\int\limits_{-\tau_L}^{\tau_L}
   {{\rm d}\tau\over\tau}\,W_k(\tau-i\epsilon) \,,
\end{equation}
where $\tau_L=\Lambda_{\rm V}^2/m_\tau^2$ is the position of the
Landau pole in the coupling constant $a(\tau m_\tau^2)$ defined in
(\ref{adef}). The functions $W_k(\tau)$ are related to weighted
integrals of the distribution function $w_D(\tau)$ \cite{part1}:
\begin{equation}
   W_k(\tau) = (k+1)\,\int\limits_0^1\!{\rm d}x\,x^k\,w_D(\tau/x) \,.
\end{equation}
It is possible to perform these integrals explicitly. For $k=0$, we
find that
\begin{eqnarray}\label{W0fun}
   W_0(\tau) &=& {32\tau\over 3}\,\Bigg\{ 4 - 3\zeta(3)
    - {15\over 4}\,\tau - 2 L_3(-\tau) + \Big[ 2\tau + L_2(-\tau)
    \Big]\,\ln\tau \nonumber\\
   &&\qquad\mbox{}- (1+\tau)\,\Big[ L_2(-\tau) + \ln\tau\,
    \ln(1+\tau) \Big] \Bigg\} \,;\quad \tau<1 \,, \nonumber\\
   && \\
   W_0(\tau) &=& {32\tau\over 3}\,\Bigg\{ - 1 + {5\over 4\tau}
    + 2 L_3(-\tau^{-1}) - \bigg( 1 - {1\over 2\tau}
    - L_2(-\tau^{-1}) \bigg)\,\ln\tau \nonumber\\
   &&\qquad\mbox{}- (1+\tau)\,\Big[ L_2(-\tau^{-1}) - \ln\tau\,
    \ln(1+\tau^{-1}) \Big] \Bigg\} \,;\quad \tau>1 \,. \nonumber
\end{eqnarray}
$L_3(x)=\int_0^x\frac{{\rm d}t}{t}\,L_2(t)$ is the trilogarithm. For
$k=1$ we obtain
\begin{eqnarray}
   W_1(\tau) &=& {64\tau\over 3}\,\Bigg\{ \bigg( {7\over 2} -
3\zeta(3)
    - 2 L_3(-\tau) \bigg)\,\tau - \bigg( {7\over 4} - L_2(-\tau)
    \bigg)\,\tau\,\ln\tau \nonumber\\
   &&\qquad\mbox{}+ (1+\tau)\,\Big[ L_2(-\tau) + \ln\tau\,
    \ln(1+\tau) \Big] \Bigg\} \,;\quad \tau<1 \,, \nonumber\\
   && \\
   W_1(\tau) &=& {64\tau\over 3}\,\Bigg\{ 3 + {1\over 2\tau}
    + 2\tau\,L_3(-\tau^{-1}) + \bigg( 2 + {1\over 4\tau}
    +\tau\,L_2(-\tau^{-1}) \bigg)\,\ln\tau \nonumber\\
   &&\qquad\mbox{}+ (1+\tau)\,\Big[ L_2(-\tau^{-1}) - \ln\tau\,
    \ln(1+\tau^{-1}) \Big] \Bigg\} \,;\quad \tau>1 \,. \nonumber
\end{eqnarray}
For the remaining cases, $k\ge 2$, we find
\begin{eqnarray}\label{Wkfuns}
   W_k(\tau) &=& {32\tau\over 3}\,(k+1)\,\Bigg\{ \bigg(
    {k^2 +k+1\over k^3} + {3 k+5\over 4(k+1)^2}
    - {7 k^2 -18 k+15\over 4(k-1)^3}
    - {2 (-1)^k\over k(k-1)}\,S_2^{(k)}(-1) \bigg)\,\tau^k
    \nonumber\\
   &&\qquad\mbox{}+ {7 k^2 -18 k+15\over 4(k-1)^3}\,\tau
    - {k-2\over(k-1)^2}\,\tau\,\ln\tau \nonumber\\
   &&\qquad\mbox{}+ \bigg( {1\over k} + {\tau\over k-1}
    + {(-\tau)^k\over k(k-1)} \bigg)\,\Big[ L_2(-\tau)
    + \ln\tau\,\ln(1+\tau) \Big] \nonumber\\
   &&\qquad\mbox{}+ {(-\tau)^k\over k(k-1)}\,\bigg( S_2^{(k)}(-\tau)
    + S_1^{(k)}(-\tau)\,\ln\tau - {1\over 2}\,\ln^2\!\tau \bigg)
    \Bigg\} \,;\quad \tau<1 \,, \nonumber\\
   &&\\
   W_k(\tau) &=& {32\tau\over 3}\,(k+1)\,\Bigg\{ \bigg(
    {k^2 +k+1\over k^3} + {3 k+5\over 4(k+1)^2}\,{1\over\tau}
    + \bigg( {k+1\over k^2} + {1\over 2(k+1)}\,{1\over\tau}
    \bigg)\,\ln\tau \nonumber\\
   &&\qquad\mbox{}+ \bigg( {1\over k} + {\tau\over k-1}
    + {(-\tau)^k\over k(k-1)} \bigg)\,\Big[ L_2(-\tau^{-1})
    - \ln\tau\,\ln(1+\tau^{-1}) \Big] \nonumber\\
   &&\qquad\mbox{}- {(-\tau)^k\over k(k-1)}\,\bigg( S_2^{(k)}(-\tau)
    + S_1^{(k)}(-\tau)\,\ln\tau \bigg) \Bigg\} \,;\quad \tau>1 \,,
    \nonumber
\end{eqnarray}
with
\begin{equation}
   S_n^{(k)}(x) = \sum_{l=1}^{k-1}\,{x^{-l}\over l^n} \,.
\end{equation}
These functions obey the normalization
\begin{equation}
   \int\limits_0^\infty {{\rm d}\tau\over\tau}\,W_k(\tau) = 4 \,,
\end{equation}
which yields the unit coefficient in front of $\alpha_s/\pi$ in the
perturbative series for $M_k$. Moreover, we note that
$W_\infty(\tau)=w_D(\tau)$. In Fig.~\ref{fig:3}, we show $W_k(\tau)$
as a function of $\ln\tau$ for a few values of $k$. For $k\to\infty$,
these functions converge towards the function $w_D(\tau)$ depicted in
Fig.~\ref{fig:1}.

Weighted integrals of the functions $W_k(\tau)$ with powers of
$\ln\tau$ determine the coefficients in the perturbative expansion of
the moments $M_k$ in the large-$\beta_0$ limit. Expanding
(\ref{MkBorel}) to order $\alpha_s^3$, one finds \cite{part1}
\begin{equation}
   M_k^{{\rm large}-\beta_0} = 1 + {\alpha_s\over\pi}
   - {\beta_0\over 4}\,(\langle\ln\tau\rangle + C)\,\bigg(
   {\alpha_s\over\pi} \bigg)^2 + {\beta_0^2\over 16}\,
   \bigg( \langle\ln^2\!\tau\rangle + 2 C\,
   \langle\ln\tau\rangle + C^2 - {\pi^2\over 3} \bigg)
   \bigg( {\alpha_s\over\pi} \bigg)^3 + O(\alpha_s^4) \,,
\end{equation}
where $\alpha_s=\alpha_s(m_\tau^2)$, $C=-5/3$ for the
$\overline{\rm MS}$ scheme ($C=0$ for the V scheme), and
\begin{eqnarray}
   \langle\ln\tau\rangle &=& {1\over 4}\,\int\limits_0^\infty
    {{\rm d}\tau\over\tau}\,\ln\tau\,W_k(\tau) = 4\zeta(3)
    - {23\over 6} - {1\over k+1} \,, \nonumber\\
   \langle\ln^2\!\tau\rangle &=& {1\over 4}\,\int\limits_0^\infty
    {{\rm d}\tau\over\tau}\,\ln^2\!\tau\,W_k(\tau) = 18 - 12\zeta(3)
    + \bigg( {23\over 3} - 8\zeta(3) \bigg)\,{1\over k+1}
    + {2\over(k+1)^2} \,.
\end{eqnarray}
This does indeed reproduce the result given in (\ref{largeb0}).

In Table~\ref{tab:4}, we show the resummed results for the first six
moments as a function of the value of the coupling constant
$\alpha_s(m_\tau^2)$ in the $\overline{\rm MS}$ scheme, which
according to (\ref{LamV}) defines the scale parameter $\Lambda_{\rm
V}$. In order to perform the numerical integrations in
(\ref{MkBorel}), it is convenient to use the asymptotic behaviour of
the functions $W_k(\tau)$ for large and small values of $\tau$, which
is given in Appendix~\ref{app:asy}. When expanded in powers of the
coupling constant, our resummation reproduces the large-$\beta_0$
limit for the perturbative coefficients given in (\ref{largeb0}). We
can correct for the missing pieces, which are known up to order
$\alpha_s^3$, using (\ref{correct}). This leads to the numbers given
in the lower portion of the table.\footnote{We note, however, that
this correction depends on the renormalization scheme. The results
presented here refer to the $\overline{\rm MS}$ scheme.}

\begin{table}[t]
\centerline{\parbox{15cm}{\caption{\label{tab:4}\protect\small\sc
Predictions for the resummed perturbative contributions to the
moments obtained from the principle value of the Borel integral. The
lower portion of the table contains the results corrected for the
exact coefficients up to order $\alpha_s^3$.}}}
\vspace{0.5cm}
\centerline{\begin{tabular}{c|cccccc}
\hline\hline
\rule[-0.2cm]{0cm}{0.7cm} $\alpha_s(m_\tau^2)$ & $M_0^{\rm Borel}$ &
 $M_1^{\rm Borel}$ & $M_2^{\rm Borel}$ & $M_3^{\rm Borel}$ &
 $M_4^{\rm Borel}$ & $M_5^{\rm Borel}$ \\
\hline
0.26 & 1.122 & 1.097 & 1.094 & 1.092 & 1.092 & 1.091 \\
0.28 & 1.136 & 1.103 & 1.100 & 1.099 & 1.098 & 1.097 \\
0.30 & 1.151 & 1.108 & 1.105 & 1.104 & 1.103 & 1.103 \\
0.32 & 1.168 & 1.113 & 1.108 & 1.108 & 1.108 & 1.107 \\
0.34 & 1.186 & 1.117 & 1.109 & 1.110 & 1.111 & 1.111 \\
0.36 & 1.205 & 1.121 & 1.109 & 1.111 & 1.112 & 1.112 \\
0.38 & 1.226 & 1.126 & 1.105 & 1.109 & 1.111 & 1.112 \\
\hline
0.26 & 1.119 & 1.093 & 1.090 & 1.088 & 1.087 & 1.087 \\
0.28 & 1.133 & 1.098 & 1.095 & 1.093 & 1.092 & 1.092 \\
0.30 & 1.148 & 1.103 & 1.098 & 1.097 & 1.097 & 1.096 \\
0.32 & 1.163 & 1.106 & 1.101 & 1.100 & 1.100 & 1.099 \\
0.34 & 1.180 & 1.109 & 1.100 & 1.101 & 1.101 & 1.101 \\
0.36 & 1.199 & 1.112 & 1.098 & 1.100 & 1.100 & 1.101 \\
0.38 & 1.218 & 1.114 & 1.093 & 1.095 & 1.097 & 1.098 \\
\hline\hline
\end{tabular}}
\vspace{0.5cm}
\end{table}

The renormalon ambiguity in the perturbative contribution to the
moments $M_k$ can be obtained by inspecting the corresponding Borel
transforms
\begin{equation}
   \widehat S_k(u) = {k+1\over k+1-u}\,{\sin\pi u\over\pi u}\,
   \widehat S_D(u) \,,
\end{equation}
or by performing the integrals \cite{BBnew}
\begin{equation}
   \Delta M_k^{\rm ren} = {1\over\pi\beta_0}\,\mbox{Im}\,
   \int\limits_{-\tau_L}^0 {{\rm d}\tau\over\tau}\,
   W_k(\tau-i\epsilon) \,.
\end{equation}
The structure of IR renormalon poles is as follows: For $k=0$
there are single poles at $u=3,4,\dots$; for $k=1$ there are single
poles at $u=2,3,\dots$; for $k=2$ there is a double pole at $u=3$ and
singles poles at $u=4,5,\dots$; for $k\ge 2$ there is a double pole
at $u=k+1$ and single poles at $u\ge 3$ ($u\ne k+1$). For the
corresponding renormalon ambiguities, we obtain
\begin{eqnarray}\label{Mkren}
   \Delta M_k^{\rm ren} &=& - {16\over 9\beta_0}\,{k+1\over k-2}\,
    {\Lambda_{\rm V}^6\over m_\tau^6} + O(m_\tau^{-8})
    \,;\quad k\ne 1,2 \,, \nonumber\\
   \Delta M_1^{\rm ren} &=& - {8\over\beta_0}\,
    {\Lambda_{\rm V}^4\over m_\tau^4} + {32\over 9\beta_0}\,
    {\Lambda_{\rm V}^6\over m_\tau^6} + O(m_\tau^{-8}) \,,
\nonumber\\
   \Delta M_2^{\rm ren} &=& -{16\over 3\beta_0}\,\bigg(
    \ln{m_\tau^2\over\Lambda_{\rm V}^2} + {11\over 6} \bigg)\,
    {\Lambda_{\rm V}^6\over m_\tau^6} + O(m_\tau^{-8}) \,.
\end{eqnarray}

\begin{table}[htb]
\centerline{\parbox{15cm}{\caption{\label{tab:dn}\protect\small\sc
Perturbative coefficients for the first three moments in the
large-$\beta_0$ limit.}}}
\vspace{0.5cm}
\centerline{\begin{tabular}{cc|rrrrr}
\hline\hline
\rule[-0.2cm]{0cm}{0.7cm} & $n$ & 1 & 2~~~~ & 3~~~~ & 4~~~~ &
 5~~~~ \\
\hline
\rule[-0.1cm]{0cm}{0.6cm}
$k=0$ & $\delta_n^{\overline{\rm MS}}\,\beta_0^{n-1}$ & 1 &
 3.80649 & 16.1854 & 56.3140 & 223.935 \\
\rule[-0.1cm]{0cm}{0.6cm}
 & $\delta_n^{\rm V}\,\beta_0^{n-1}$ & 1 & 0.05649 & 1.69928 &
 $-17.9203$ & 139.694 \\
\hline
\rule[-0.1cm]{0cm}{0.6cm}
$k=1$ & $\delta_n^{\overline{\rm MS}}\,\beta_0^{n-1}$ & 1 &
 2.68149 & 5.08959 & $-35.7603$ & $-443.813$ \\
\rule[-0.1cm]{0cm}{0.6cm}
 & $\delta_n^{\rm V}\,\beta_0^{n-1}$ & 1 & $-1.06851$ & $-0.95907$ &
 $-32.6273$ & 154.154 \\
\hline
\rule[-0.1cm]{0cm}{0.6cm}
$k=2$ & $\delta_n^{\overline{\rm MS}}\,\beta_0^{n-1}$ & 1 &
 2.30649 & 2.51598 & $-47.2767$ & $-424.721$ \\
\rule[-0.1cm]{0cm}{0.6cm}
 & $\delta_n^{\rm V}\,\beta_0^{n-1}$ & 1 & $-1.44351$ & $-0.72018$ &
 $-31.0109$ & 207.944 \\
\hline\hline
\end{tabular}}
\vspace{0.5cm}
\end{table}

Let us again study the behaviour of the perturbative series in the
large-$\beta_0$ limit. The asymptotic growth of the expansion
coefficients $\delta_n$ in (\ref{kappan}), which can
be obtained by expanding the Borel transform $\widehat S_k(u)$ in
powers of $u$, is determined by the UV
renormalon at $u=-1$. However, the behaviour in low orders is
governed by a complicated interplay of this UV renormalon with the
nearest IR renormalon, which is located at $u=2$ for $k=1$ and $u=3$
for $k\ne 1$. For the first three moments, we show the coefficients
$\delta_n$ in Table~\ref{tab:dn}. In Table~\ref{tab:momser}, we show
the partial sums $M_k^{(N)} = 1 +
\sum_{n=1}^N\,\delta_n\,\beta_0^{n-1}\, (\alpha_s/\pi)^n$ and compare
them to the asymptotic value obtained from the resummation of
renormalon chains. As before, the minimal term in the series is
reached later in the $\overline{\rm MS}$ scheme than in the V scheme.
If one truncates the series at the minimal term, in both schemes the
differences to the resummed results are typically of order $10^{-3}$
for $k=0$, and of order 1--2\% for $k=1,2$. In all cases these
differences are much larger than the renormalon ambiguities.

\begin{table}[htb]
\centerline{\parbox{15cm}{\caption{\label{tab:momser}
\protect\small\sc
Partial sums of the perturbative series for the first three moments
in the large-$\beta_0$ limit.}}}
\vspace{0.5cm}
\centerline{\begin{tabular}{c|cc|cc|cc}
\hline\hline
\rule[-0.2cm]{0cm}{0.7cm} $N$ &
 $M_{0,{\overline{\rm MS}}}^{(N)}$ & $M_{0,{\rm V}}^{(N)}$ &
 $M_{1,{\overline{\rm MS}}}^{(N)}$ & $M_{1,{\rm V}}^{(N)}$ &
 $M_{2,{\overline{\rm MS}}}^{(N)}$ & $M_{2,{\rm V}}^{(N)}$ \\
\hline
1 & 1.1019 & 1.1648 & 1.1019 & 1.1648 & 1.1019 & 1.1648 \\
2 & 1.1414 & 1.1663 & 1.1297 & 1.1358 & 1.1258 & 1.1256 \\
3 & 1.1585 & 1.1740 & 1.1351 & 1.1315 & 1.1284 & 1.1224 \\
4 & 1.1645 & 1.1607 & 1.1312 & 1.1074 & 1.1234 & 1.0995 \\
5 & 1.1670 & 1.1777 & 1.1263 & 1.1262 & 1.1187 & 1.1248 \\
$\infty$ & \multicolumn{2}{c|}{$1.1678\pm .00003$} &
 \multicolumn{2}{c|}{$1.1131\pm 0.0039$} &
 \multicolumn{2}{c}{$1.1080\pm 0.0008$} \\
\hline\hline
\end{tabular}}
\vspace{0.5cm}
\end{table}

An interesting question that can be addressed using the
large-$\beta_0$ limit is whether the resummation procedure of
Le~Diberder and Pich \cite{LP1} improves the convergence of the
perturbative series. To answer this question, we investigate the
partial sums
\begin{equation}
   M_k^{{\rm LP},(N)} = 1 + \sum_{n=1}^N\,\kappa_n\,
   \beta_0^{n-1}\,I_n^{k+1}(m_\tau) \,,
\end{equation}
where the integrals $I_n^{k+1}(m_\tau)$ are evaluate using the
one-loop $\beta$-function for the running coupling constant (see
Appendix~\ref{app:int}). This ensures that for $N\to\infty$ one
recovers the large-$\beta_0$ limit of the series. The partial sums
for the first three moments are shown in Table~\ref{tab:momLP}. The
results are interesting, as they indicate that the convergence is not
improved in a significant way. In some cases (such as $k=0$),
fixed-order perturbation theory even converges better towards the
asymptotic result. We believe that this observation is not specific
for the large-$\beta_0$ limit, since the large $\pi^2$-terms resummed
in the approach of Le~Diberder and Pich are retained in this limit.

\begin{table}[htb]
\centerline{\parbox{15cm}{\caption{\label{tab:momLP}
\protect\small\sc
Partial sums of the perturbative series for the first three moments
obtained using the large-$\beta_0$ limit of the resummation procedure
of Le~Diberder and Pich.}}}
\vspace{0.5cm}
\centerline{\begin{tabular}{c|cc|cc|cc}
\hline\hline
\rule[-0.2cm]{0cm}{0.7cm} $N$ &
 $M_{0,{\overline{\rm MS}}}^{{\rm LP},(N)}$ &
 $M_{0,{\rm V}}^{{\rm LP},(N)}$ &
 $M_{1,{\overline{\rm MS}}}^{{\rm LP},(N)}$ &
 $M_{1,{\rm V}}^{{\rm LP},(N)}$ &
 $M_{2,{\overline{\rm MS}}}^{{\rm LP},(N)}$ &
 $M_{2,{\rm V}}^{{\rm LP},(N)}$ \\
\hline
1 & 1.1155 & 1.1852 & 1.0973 & 1.1350 & 1.0940 & 1.1316 \\
2 & 1.1343 & 1.1235 & 1.1098 & 1.1094 & 1.1059 & 1.1031 \\
3 & 1.1526 & 1.1903 & 1.1180 & 1.1104 & 1.1141 & 1.1208 \\
4 & 1.1552 & 1.1309 & 1.1185 & 1.1415 & 1.1148 & 1.1103 \\
5 & 1.1622 & 1.2018 & 1.1175 & 1.0503 & 1.1155 & 1.1341 \\
$\infty$ & \multicolumn{2}{c|}{$1.1678\pm .00003$} &
 \multicolumn{2}{c|}{$1.1131\pm 0.0039$} &
 \multicolumn{2}{c}{$1.1080\pm 0.0008$} \\
\hline\hline
\end{tabular}}
\vspace{0.5cm}
\end{table}

To summarize, in the first part of this paper we have investigated
three schemes to calculate the perturbative contributions to the
spectral moments $M_k$: fixed-order perturbation theory, the
resummation procedure of Le~Diberder and Pich, and resummed
perturbation theory in the large-$\beta_0$ limit. The results are
summarized in Tables~\ref{tab:1}, \ref{tab:2} and \ref{tab:4}. As
there is no strong argument to prefer one of these schemes over the
others, the differences between the numerical results must be
considered as theoretical uncertainties in the perturbative
calculation of the moments. These differences are of order a few per
cent. We note that taking ratios of moments (as proposed in
Ref.~\cite{LP2}) does not improve the accuracy, although such ratios
are known exactly to order $\alpha_s^4$. For instance, the
predictions for the ratio $M_1/M_0$ differ by a larger amount than
the predictions for the individual moments. In the second part of the
paper, we shall investigate what this uncertainty implies for the
sensitivity to power corrections.

\section{Power corrections}
\label{sec:5}

One of the goals of analysing moments of the $\tau$ hadronic spectrum
is to test QCD at the level of nonperturbative effects, which
manifest themselves in the form of power corrections \cite{LP2}. In
the standard approach of SVZ, these corrections are parametrized in
terms of vacuum expectation values of local, gauge-invariant
operators, the so-called condensates \cite{SVZ}. There have been
attempts to extract the condensates of dimension four, six and even
eight from the analysis of hadronic $\tau$ decays
\cite{Laun}--\cite{GiBo}, \cite{ALEPH,CLEO}. However, the feasibility
of such determinations is limited by the fact that, in order to be
sensitive to power-suppressed effects, one has to control
perturbation theory to a sufficient level of accuracy.

There is also a conceptual need for an analysis of power corrections.
As we have seen, the perturbative definition of the moments is
ambiguous because of the presence of IR renormalons. In order to deal
with these ambiguities in a consistent way, one is forced to add
other nonperturbative contributions. Hence, we shall write
\begin{equation}
   {\cal M}_k^{(J)} = {\cal M}_{k,{\rm pert}}^{(J)}
   + {\cal M}_{k,{\rm power}}^{(J)} = M_k\,\delta_{J=1}
   + {\cal M}_{k,{\rm power}}^{(J)} \,.
\end{equation}
Only the sum of the perturbative and nonperturbative contributions
can be expected to be well defined and unambiguous. We shall now
present a model-independent analysis of power corrections, which does
not rely on assumptions about their origin. Our goal is to relate the
power corrections to the moments to those of the euclidean
correlators $D^{(J)}(Q^2)$ defined in (\ref{Dijdef}), for which we
write
\begin{equation}\label{Dpower}
   D^{(J)}(Q^2) = D_{\rm pert}^{(J)}(Q^2) + D_{\rm power}^{(J)}(Q^2)
   = D(Q^2)\,\delta_{J=1} + \sum_{n=1}^\infty\,
   {\langle O_{2n}^{(J)}(Q^2)\rangle\over Q^{2n}} \,.
\end{equation}
The nonperturbative quantities $\langle O_{2n}^{(J)}(Q^2)\rangle$
scale like $\Lambda^{2n}$. The only assumption we shall make is that
these quantities have a weak, logarithmic dependence on $Q^2$.
Writing then
\begin{equation}\label{nabla}
   \langle O_{2n}^{(J)}(Q^2)\rangle = \exp\bigg(
   \ln{Q^2\over m_\tau^2}\,{{\rm d}\over{\rm d}\ln m_\tau^2}
   \bigg)\,\langle O_{2n}^{(J)}(m_\tau^2)\rangle
   \equiv \bigg( {Q^2\over m_\tau^2} \bigg)^{\!\nabla}\,
   \langle O_{2n}(m_\tau^2)\rangle \,,
\end{equation}
it is justified to treat $\nabla={\rm d}/{\rm d}\ln m_\tau^2$ as a
small parameter. In fact, in QCD the nonperturbative quantities
$\langle O_{2n}^{(J)}(m_\tau^2)\rangle$ depend on $m_\tau$ through
the running coupling constant and through the running quark
masses,\footnote{This becomes explicit if these quantities are
expressed in terms of scale-invariant condensates, see
Appendix~\protect\ref{app:QCD}.}
so that
\begin{eqnarray}
   \nabla &=& \gamma_m[\alpha_s(m_\tau^2)]\,m_q^2\,
    {\partial\over\partial m_q^2} - \beta[\alpha_s(m_\tau^2)]\,
    \alpha_s\,{\partial\over\partial\alpha_s} \nonumber\\
   &\simeq& - {\alpha_s(m_\tau^2)\over\pi}\,\bigg(
    2 m_q^2\,{\partial\over\partial m_q^2}
    + {9\over 4}\,\alpha_s\,{\partial\over\partial\alpha_s}
    \bigg) \,.
\end{eqnarray}
Here $\gamma_m=-2\alpha_s/\pi+\dots$ is the anomalous dimension of
the running quark mass. Inserting expression (\ref{nabla}) into the
contour integral (\ref{Mcirc}), we find
\begin{equation}
   {\cal M}_{k,{\rm power}}^{(J)} = \sum_{n=1}^\infty\,
   {1\over m_\tau^{2n}}\,I_{k,n}(\nabla)\,
   \langle O_{2n}^{(J)}(m_\tau^2)\rangle \,,
\end{equation}
with
\begin{eqnarray}
   I_{k,n}(\nabla) &=& {1\over 2\pi i}\,\oint\limits_{|x|=1}
    {{\rm d}x\over x}\,(1 - x^{k+1})\,(-x)^{\nabla-n} \nonumber\\
   &=& (-1)^{n+1}\,{k+1\over(k+1-n+\nabla)(n-\nabla)}\,
    {\sin(\pi\nabla)\over\pi} \,.
\end{eqnarray}
To second order in $\nabla$, we obtain
\begin{equation}
   I_{k,k+1}(\nabla) = (-1)^k\,\bigg\{ 1 + {\nabla\over k+1}
   + \bigg( {1\over(k+1)^2} - {\pi^2\over 6} \bigg)\,\nabla^2
   + O(\nabla^3) \bigg\}
\end{equation}
for $n=k+1$, and
\begin{equation}
   I_{k,n}(\nabla) = (-1)^{n+1}\,{k+1\over(k+1-n)n}\,\bigg\{
   \nabla + {k+1-2 n\over(k+1-n)n}\,\nabla^2
   + O(\nabla^3) \bigg\}
\end{equation}
for $n\ne k+1$. Hence, if the logarithmic dependence of $\langle
O_{2n}(m_\tau^2)\rangle$ on $m_\tau$ is neglected, the $k$-th moment
projects out the power corrections of dimension $d=2k+2$:
\begin{equation}
   {\cal M}_{k,{\rm power}}^{(J)} = (-1)^k\,
   {\langle O_{2k+2}^{(J)}(m_\tau^2)\rangle\over m_\tau^{2k+2}}
   + O(\nabla) \,.
\end{equation}
The most important effect of the scale dependence is to induce
contributions from lower-dimensional operators.

Using the above results we can derive explicit expressions for the
power corrections to the moments, treating both $\nabla$ and
$1/m_\tau^2$ as small parameters. Neglecting terms of order
$\nabla^3\,\langle O_2\rangle/m_\tau^2$, $\nabla^2\,\langle
O_4\rangle/m_\tau^4$, $\nabla\,\langle O_6\rangle/m_\tau^6$ and
operators of dimension larger or equal to eight, we find
\begin{eqnarray}\label{Mkpower}
   {\cal M}_{0,{\rm power}}^{(0+1)}
   &=& \bigg[ 1 + \nabla + \bigg( 1 - {\pi^2\over 6} \bigg)\,
    \nabla^2 \bigg]\,{\langle O_2^{(0+1)}\rangle\over m_\tau^2}
    + {\nabla\over 2}\,{\langle O_4^{(0+1)}\rangle\over m_\tau^4}
    + \dots \,, \nonumber\\
   {\cal M}_{1,{\rm power}}^{(J)}
   &=& 2\nabla\,{\langle O_2^{(J)}\rangle\over m_\tau^2}
    - \bigg( 1 + {\nabla\over 2} \bigg)\,
    {\langle O_4^{(J)}\rangle\over m_\tau^4} + \dots
    \,, \nonumber\\
   {\cal M}_{2,{\rm power}}^{(J)}
   &=& \bigg( {3\over 2}\,\nabla + {3\over 4}\,\nabla^2 \bigg)\,
    {\langle O_2^{(J)}\rangle\over m_\tau^2}
    - {3\over 2}\,\nabla\,{\langle O_4^{(J)}\rangle\over m_\tau^4}
    + {\langle O_6^{(J)}\rangle\over m_\tau^6} + \dots
    \,, \nonumber\\
   {\cal M}_{k,{\rm power}}^{(J)}
   &=& \bigg( {k+1\over k}\,\nabla + {k^2-1\over k^2}\,\nabla^2
    \bigg)\,{\langle O_2^{(J)}\rangle\over m_\tau^2}
    - {k+1\over 2(k-1)}\,\nabla\,
    {\langle O_4^{(J)}\rangle\over m_\tau^4} + \dots \,;\quad
    k\ge 3 \,.
\end{eqnarray}
Thus, the power corrections are parametrized by a set of fundamental
nonperturbative parameters $\nabla^k\,\langle O_{2n}^{(J)}\rangle$.

We now apply this general formalism to calculate the leading power
corrections in the SVZ approach \cite{SVZ}. The corresponding
corrections $\langle O_{2n}^{(J)}\rangle$ to the current correlators
$D^{(J)}$ have been calculated, at next-to-leading order in the
coupling constant, by several authors. They have been summarized and
rewritten in terms of scale-invariant condensates in
Ref.~\cite{Rtau1}. For the corrections of dimension two and four, the
results are \cite{BNRY}--\cite{CGS}
\begin{eqnarray}\label{O2nexpr}
   \langle O_2^{(0+1)}\rangle &=& -3\,\bigg( 1 + {13\over 3}\,
    {\alpha_s\over\pi} \bigg)\,|\,V_{uj}|^2\,(m_u^2+m_j^2) \,,
    \nonumber\\
   \langle O_4^{(0+1)}\rangle &=& {2\pi^2\over 3}\,
    \bigg( 1 - {11\over 18}\,{\alpha_s\over\pi} \bigg)\,
    \bigg\langle {\alpha_s\over\pi}\,G^2 \bigg\rangle \nonumber\\
   &&\mbox{}+ 8\pi^2\,\bigg( 1 - {\alpha_s\over\pi} \bigg)\,
    |\,V_{uj}|^2\,\langle m_u\bar\psi_u\psi_u
    + m_j\bar\psi_j\psi_j\rangle + {32\pi^2\over 27}\,
    {\alpha_s\over\pi}\,\sum_k\,\langle m_k\bar\psi_k\psi_k \rangle
    \nonumber\\
   &&\mbox{}+ 12\,|\,V_{uj}|^2\,m_u^2 m_j^2
    - {2\over 7}\,\sum_k\,m_k^4 - \bigg( {24\over 7}\,
    {\pi\over\alpha_s} + 1 \bigg)\,|\,V_{uj}|^2\,(m_u^4 + m_j^4)
    \,, \nonumber\\
   \langle O_2^{(0)}\rangle &=& 6\,\bigg( {\pi\over\alpha_s}
    - {11\over 4} \bigg)\,|\,V_{uj}|^2\,(m_u^2+m_j^2)
    + \mbox{const.} \,, \nonumber\\
   \langle O_4^{(0)}\rangle &=& 8\pi^2\,|\,V_{uj}|^2\,
    \langle m_u\bar\psi_u\psi_u + m_j\bar\psi_j\psi_j\rangle
    + 12\,|\,V_{uj}|^2\,m_u^2 m_j^2 \nonumber\\
   &&\mbox{}- \bigg( {24\over 7}\,{\pi\over\alpha_s} + {10\over 7}
    \bigg)\,|\,V_{uj}|^2\,(m_u^4 + m_j^4) \,.
\end{eqnarray}
Here $\alpha_s=\alpha_s(m_\tau^2)$ and $m_q=m_q(m_\tau^2)$. A
summation over $j=d,s$ is understood. The terms proportional to
$1/\alpha_s$ arises from IR logarithms, which can be resummed into
the running coupling constant by means of the RGE \cite{Gene}. Note
that $\langle O_2^{(0)}\rangle$ is defined only up to a constant
(with respect to $m_\tau^2$), which depends on the renormalization
scheme. However, only derivatives of $\langle O_2^{(0)}\rangle$
appear in physical quantities.

The most important power corrections of dimension six come from
four-quark operators; the coefficient of the operator $G^3$ vanishes
to leading order in $\alpha_s$ \cite{Hub,Dubo}. Operators containing
powers of the light quark masses are strongly suppressed and can
safely be neglected. The coefficient functions for the four-quark
operators were calculated to next-to-leading order in
Ref.~\cite{Lani}. In the chiral limit there are no contributions to
the $J=0$ correlator, and the contributions to the $J=1$ correlator
are\footnote{We have rewritten the result given in
Ref.~\protect\cite{Lani} using Fierz identities. Note that, in the
notation of Ref.~\protect\cite{LP2}, one has $\langle
O_6^{(0+1)}\rangle=6\pi^2\,O(6)$.}
\begin{eqnarray}\label{4quark}
   \langle O_6^{(0+1)}\rangle &=& -{64\pi^4\over 3}\,\bigg[
    1 + \bigg( {9\over 2} - 2 L \bigg)\,{\alpha_s(\mu^2)\over\pi}
    \bigg]\,{\alpha_s(\mu^2)\over\pi}\,
    ( O_{ij}^{V1} + O_{ij}^{A1} ) \nonumber\\
   &&\mbox{}+ 16\pi^4\,\bigg[ 1 + \bigg( {87\over 16} + {1\over 4}\,
    L \bigg)\,{\alpha_s(\mu^2)\over\pi}\bigg]\,
    {\alpha_s(\mu^2)\over\pi}\,
    ( O_{ij}^{V8} + O_{ij}^{A8} ) \nonumber\\
   &&\mbox{}- {32\pi^4\over 3}\,\bigg[
    1 + \bigg( {1153\over 432} - {95\over 72}\,L \bigg)\,
    {\alpha_s(\mu^2)\over\pi} \bigg]\,{\alpha_s(\mu^2)\over\pi}\,
    \sum_k\,( O_{ik}^{V8} + O_{jk}^{V8} ) \nonumber\\
   &&\mbox{}- {16\pi^4\over 3}\,\bigg( 1 - {2\over 3}\,L \bigg)
    \bigg( {\alpha_s(\mu^2)\over\pi} \bigg)^2\,
    \sum_k\,\bigg[ O_{ik}^{A1} + O_{jk}^{A1}
    + {15\over 8}\,( O_{ik}^{A8} + O_{jk}^{A8} ) \bigg] \nonumber\\
   &&\mbox{}- {16\pi^4\over 27}\,( 1 - 6 L )
    \bigg( {\alpha_s(\mu^2)\over\pi} \bigg)^2\,
    \sum_{k,l}\,O_{kl}^{V8} \,.
\end{eqnarray}
The dependence on $m_\tau^2$ resides in $L=\ln m_\tau^2/\mu^2$, and
$\mu$ is an arbitrary renormalization scale. Since we work in the
chiral limit, it is not necessary to specify the flavour labels;
however, it is important that $i\ne j$. The four-quark operators,
which are renormalized in the $\overline{\rm MS}$ scheme, are defined
as
\begin{eqnarray}
   O_{ij}^{V1} &=& \langle \bar\psi_i\gamma_\mu\psi_i\,
    \bar\psi_j\gamma^\mu\psi_j(\mu^2) \rangle \,, \nonumber\\
   O_{ij}^{A1} &=& \langle \bar\psi_i\gamma_\mu\gamma_5\psi_i\,
    \bar\psi_j\gamma^\mu\gamma_5\psi_j(\mu^2) \rangle \,, \nonumber\\
   O_{ij}^{V8} &=& \langle \bar\psi_i\gamma_\mu T_a\psi_i\,
    \bar\psi_j\gamma^\mu T_a\psi_j(\mu^2) \rangle \,, \nonumber\\
   O_{ij}^{A8} &=& \langle \bar\psi_i\gamma_\mu\gamma_5 T_a\psi_i\,
    \bar\psi_j\gamma^\mu\gamma_5 T_a\psi_j(\mu^2) \rangle \,.
\end{eqnarray}
Since there exist no reliable estimates for these condensates, the
traditional approach is to apply the vacuum saturation (or
factorization) approximation \cite{SVZ} to relate them to the quark
condensate. Unfortunately, this approximation is inconsistent with
renormalization-group (RG) invariance \cite{Tara,Jami}. Moreover,
phenomenological analyses indicate that the factorization
approximation underestimates the contribution of four-quark
operators. One introduces a fudge factor $\rho$ to compensate for
this. At leading order in $\alpha_s$, one then obtains
\begin{equation}\label{O6fact}
   \langle O_6^{(0+1)}\rangle_{\rm fact} = {256\pi^3\over 27}\,
   \rho\alpha_s\langle\bar\psi\psi\rangle^2 \,.
\end{equation}
The combination $\rho\alpha_s\langle\bar\psi\psi\rangle^2$ is treated
as a scale-invariant phenomenological parameter, whose value has been
determined to be of order 2--$6\times 10^{-4}$~GeV$^6$
\cite{Laun}--\cite{GiBo}, corresponding to $\rho\sim 3$--8 at
$\mu=1$~GeV. The fact that phenomenology requires $\rho\gg 1$ is
unsatisfactory and indicates large violations of the factorization
hypothesis.

\begin{figure}[htb]
   \vspace{0.5cm}
   \epsfysize=3cm
   \centerline{\epsffile{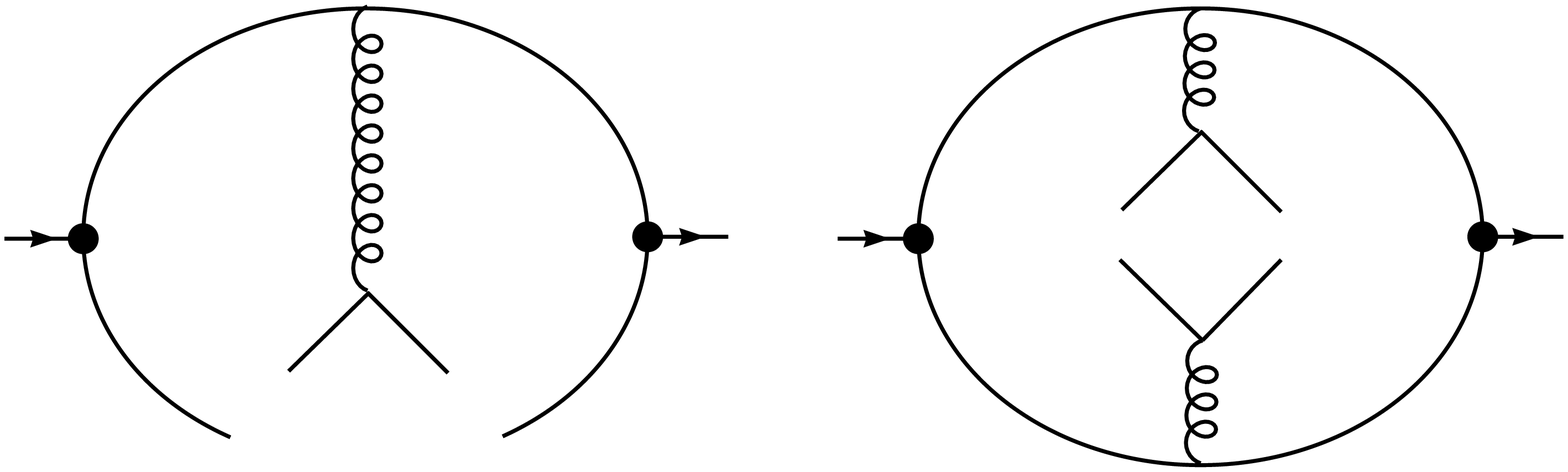}}
   \centerline{\parbox{13cm}{\caption{\label{fig:4qcond}
\protect\small\sc
Examples of leading contributions of four-quark condensates in the
large-$n_f$ limit. The first diagram is of order $n_f\alpha_s$, the
second of order $(n_f\alpha_s)^2$.}}}
\end{figure}

An interesting and, to our knowledge, new alternative to the vacuum
saturation approximation is to consider the large-$\beta_0$ limit,
which we have employed already in the calculation of the perturbative
corrections. This limit respects exact RG invariance, yet reducing
the dimension-six corrections to a single phenomenological parameter.
To obtain it, we analyse expression (\ref{4quark}) in the limit
$n_f\to\infty$. Note that there is $n_f$ dependence in the running
coupling constant as well as in the flavour sums. We find
\begin{equation}
   \langle O_6^{(0+1)}\rangle = - {64\pi^4\over 3}\,\bigg(
   {n_f\alpha_s(\mu^2)\over\pi} \bigg)^2\,O_{ij}^{V8}(\mu^2)\,
   \bigg[ {\pi\over n_f\alpha_s(\mu^2)} + {1\over 36}\,(1-6L)
   \bigg] + O(1/n_f) \,,
\end{equation}
where the product $n_f\alpha_s(\mu^2)$ is formally of order $n_f^0$.
The two terms shown in parenthesis correspond to the diagrams
depicted in Fig.~\ref{fig:4qcond}. In the next step, we replace $n_f$
by $-\frac{2}{3}\,\beta_0$ and use the one-loop expression for the
running coupling constant to find that
\begin{equation}\label{O6bubb}
   \langle O_6^{(0+1)}\rangle_{{\rm large}-\beta_0}
   = 128\pi^4\,\bigg( \ln{m_\tau^2\over\Lambda_{\rm V}^2}
   + {3\over 2} \bigg)\,a^2(\mu^2)\,O_{ij}^{V8}(\mu^2) \,,
\end{equation}
where $\Lambda_{\rm V}=e^{5/6}\,\Lambda_{\overline{\rm MS}}$, and
$a(\mu^2)$ has been defined in (\ref{adef}). The $\mu$ dependence of
$a(\mu^2)$ is exactly cancelled by the $\mu$ dependence of the
four-quark condensate \cite{Jami}. Note that the contribution
(\ref{O6bubb}) vanishes in the factorization approximation. In this
sense the above two approximations are orthogonal to each other. If
we define
\begin{equation}
   \varepsilon = {a^2(\mu^2)\,O_{ij}^{V8}(\mu^2)\over
   \langle\bar\psi\psi\rangle^2} \,,
\end{equation}
where $\langle\bar\psi\psi\rangle^2$ is evaluated at $\mu=1$~GeV, we
find that the results (\ref{O6bubb}) and (\ref{O6fact}) are of
similar magnitude if $\varepsilon\sim 1$--2\%, corresponding to
only a small deviation from vacuum saturation, which is certainly not
excluded. Hence, if one accepts that the large-$\beta_0$ limit gives
a reasonable approximation to QCD, it provides a natural explanation
of the empirical fact that the values of four-quark condensates are
much larger than predicted in the vacuum saturation approximation.

Starting from (\ref{O2nexpr}) and using the $\beta$-function and the
anomalous dimension of the running quark masses from
Appendix~\ref{app:QCD}, we find
\begin{eqnarray}
   \nabla\,\langle O_2^{(0+1)}\rangle &=& 6\,{\alpha_s\over\pi}\,
    |\,V_{uj}|^2\,(m_u^2+m_j^2) \,, \nonumber\\
   \nabla\,\langle O_4^{(0+1)}\rangle &=& \alpha_s^2\,\bigg\{
    {11\over 12}\,\bigg\langle {\alpha_s\over\pi}\,G^2 \bigg\rangle
    + 18\,|\,V_{uj}|^2\,\langle m_u\bar\psi_u\psi_u
    + m_j\bar\psi_j\psi_j\rangle - {8\over 3}\,\sum_k\,
    \langle m_k\bar\psi_k\psi_k \rangle \bigg\} \nonumber\\
   &&\mbox{}+ 6\,|\,V_{uj}|^2\,(m_u^4 + m_j^4) \,, \nonumber\\
   \nabla\,\langle O_2^{(0)}\rangle &=& {3\over 2}\,\bigg( 1
    + {23\over 3}\,{\alpha_s\over\pi} \bigg)\,|\,V_{uj}|^2\,
    (m_u^2+m_j^2) \,, \nonumber\\
   \nabla^2\,\langle O_2^{(0)}\rangle &=& -3\,{\alpha_s\over\pi}\,
    |\,V_{uj}|^2\,(m_u^2+m_j^2) \,, \nonumber\\
   \nabla\,\langle O_4^{(0)}\rangle &=& 6\,|\,V_{uj}|^2\,
    (m_u^4 + m_j^4) \,.
\end{eqnarray}
The quantity $\nabla^2\,\langle O_2^{(0+1)}\rangle$ appearing in
(\ref{Mkpower}) is of order $\alpha_s^2$ and will thus be neglected.
For completeness, we also quote the results for the derivatives of
the dimension-six operator. They are
\begin{eqnarray}
   \nabla\,\langle O_6^{(0+1)}\rangle_{\rm fact}
   &=& - {2176\pi^3\over 243}\,{\alpha_s\over\pi}\,
    \rho\alpha_s\langle\bar\psi\psi\rangle^2 \,, \nonumber\\
   \nabla\,\langle O_6^{(0+1)}\rangle_{{\rm large}-\beta_0}
   &=& 128\pi^4\,\varepsilon\langle\bar\psi\psi\rangle^2 \,.
\end{eqnarray}

To estimate the size of the various power corrections we use the
standard values of the QCD parameters, which are collected in
Appendix~\ref{app:QCD}. We find the following results for the
nonperturbative parameters corresponding to $J=0+1$:
\begin{eqnarray}\label{esti1}
   {\langle O_2^{(0+1)}\rangle\over m_\tau^2}
   &\simeq& -(1.42\pm 0.27)\times 10^{-3} \,, \nonumber\\
   {\nabla\,\langle O_2^{(0+1)}\rangle\over m_\tau^2}
   &\simeq& (0.20\pm 0.04)\times 10^{-3} \,, \nonumber\\
   {\langle O_4^{(0+1)}\rangle\over m_\tau^4}
   &\simeq& (9.2\pm 5.6)\times 10^{-3} \,, \nonumber\\
    {\nabla\,\langle O_4^{(0+1)}\rangle\over m_\tau^4}
   &\simeq& (0.18\pm 0.08)\times 10^{-3} \,, \nonumber\\
   {\langle O_6^{(0+1)}\rangle\over m_\tau^6}
   &\simeq& (3.3\pm 1.9)\times 10^{-3} \,.
\end{eqnarray}
The quoted errors reflect the uncertainty in the values of the quark
masses and vacuum condensates. We note that the values of $\langle
O_4^{(0+1)}\rangle$ and $\nabla\,\langle O_4^{(0+1)}\rangle$ are
dominated by the gluon condensate. For $J=0$, we obtain:
\begin{eqnarray}
   {\nabla\,\langle O_2^{(0)}\rangle\over m_\tau^2}
   &\simeq& (0.88\pm 0.17)\times 10^{-3} \,, \nonumber\\
   {\nabla^2\,\langle O_2^{(0)}\rangle\over m_\tau^2}
   &\simeq& -(0.10\pm 0.02)\times 10^{-3} \,, \nonumber\\
   {\langle O_4^{(0)}\rangle\over m_\tau^4}
   &\simeq& -(1.92\pm 0.72)\times 10^{-3} \,, \nonumber\\
   {\nabla\,\langle O_4^{(0)}\rangle\over m_\tau^4}
   &\simeq& 0.01\times 10^{-3} \,.
\end{eqnarray}
Most of the contributions are strongly suppressed by powers of the
small quark masses and can safely be neglected. Contributions to the
moments of order $10^{-3}$ are certainly not detectable given the
uncertainties in the perturbative calculation, and also given that
any extraction of the moments will be affected by experimental
uncertainties. Therefore, it is a safe approximation to consider the
power corrections in the chiral limit, in which case only
contributions with $J=1$ remain. The most striking prediction of the
SVZ approach is that $\langle O_2^{(1)}\rangle=0$ in the chiral
limit, since there is no gauge-invariant operator of dimension two in
QCD. As a consequence, the leading nonperturbative corrections are
induced by the gluon condensate and are of order $\langle
O_4^{(1)}\rangle/m_\tau^4\sim 1\%$. We expect that corrections of
dimension six are suppressed, relative to this, by about a factor 3,
which is beyond the precision reachable in a realistic analysis.
Therefore, we believe that the primary goals must be to measure
$\alpha_s(m_\tau^2)$ with a minimum contamination from power
corrections, to test the absence of dimension-two operators in QCD,
and to extract a value for the gluon condensate. We shall discuss
some strategies how to pursue these goals in Sect.~\ref{sec:7}.
Before, however, we will demonstrate how the inclusion of power
corrections cures the IR renormalon problem.

\section{Cancellation of renormalon ambiguities}
\label{sec:6}

The structure of power corrections simplifies greatly in the combined
large-$\beta_0$ and chiral limit. The nonperturbative parameters
remaining in this limit are the ones needed to absorb the renormalon
ambiguities in the perturbative calculations of Sects.~\ref{sec:3}
and \ref{sec:4}. We find
\begin{eqnarray}
   \langle O_4^{(1)}\rangle_{{\rm large}-\beta_0}
   &=& {2\pi^2\over 3}\,\bigg\langle {\alpha_s\over\pi}\,G^2
    \bigg\rangle \,, \nonumber\\
   \langle O_6^{(1)}\rangle_{{\rm large}-\beta_0}
   &=& 128\pi^4\,\bigg( \ln{m_\tau^2\over\Lambda_{\rm V}^2}
    + {3\over 2} \bigg)\,\varepsilon\langle\bar\psi\psi\rangle^2 \,,
    \nonumber\\
   \nabla\,\langle O_6^{(1)}\rangle_{{\rm large}-\beta_0}
   &=& 128\pi^4\,\varepsilon\langle\bar\psi\psi\rangle^2 \,.
\end{eqnarray}
All other contributions of dimension up to six vanish in this limit;
in particular all contributions for $J=0$. It is satisfying to
observe that the potentially large power corrections survive in the
large-$\beta_0$ limit. From the numerical discussion of the preceding
section, it follows that all other terms are of order $10^{-3}$ or
less (in the appropriate units of $m_\tau$).

Let us now demonstrate how the renormalon ambiguities of perturbation
theory can be cured by adding nonperturbative corrections. To start
with, consider the euclidean correlator $D^{(1)}(Q^2)$. In the
combined large-$\beta_0$ and chiral limit, we find from (\ref{Dsum})
and (\ref{Dpower})
\begin{equation}\label{Dsumtot}
   D^{(1)}(Q^2) = 1 + {1\over\beta_0}\,\int\limits_0^\infty
   {{\rm d}\tau\over\tau}\,w_D(\tau)\,a(\tau Q^2)
   + {\langle O_4^{(1)}\rangle\over Q^4}
   + {\langle O_6^{(1)}(Q^2)\rangle\over Q^6} + O(Q^{-8}) \,.
\end{equation}
The perturbative contribution is ambiguous because of the Landau pole
in the running coupling constant $a(\tau Q^2)$. The corresponding
renormalon ambiguity has been given in (\ref{Dren}). To obtain a
meaningful result, we have to require that, order by order in
$1/Q^2$, these ambiguities are cancelled by corresponding ambiguities
in the definition of the nonperturbative parameters $\langle
O_{2n}^{(1)}\rangle$. The philosophy is the following: The values of
the nonperturbative parameters become meaningful only after one
specifies a resummation prescription for the perturbative series,
i.e.\ a prescription to regulate the Landau pole in the integral over
$\tau$ in (\ref{Dsumtot}). Changing this prescription changes the
values of the nonperturbative parameters in such a way that the total
answer remains the same \cite{Muel}. Hence, to order $1/Q^6$ we have
to require that
\begin{equation}
   {8\over\beta_0}\,{\Lambda_{\rm V}^4\over Q^4}
   - {16\over 3\beta_0}\,\bigg( \ln{Q^2\over\Lambda_{\rm V}^2}
   + {3\over 2} \bigg)\,{\Lambda_{\rm V}^6\over Q^6}
   + {\Delta\langle O_4^{(1)}\rangle_{\rm ren}\over Q^4}
   + {\Delta\langle O_6^{(1)}(Q^2)\rangle_{\rm ren}\over Q^6} = 0 \,.
\end{equation}
Note that the momentum dependence of $\langle O_6^{(1)}(Q^2)\rangle$,
which is obtained from (\ref{O6bubb}) by the replacement $m_\tau^2\to
Q^2$, is precisely of the form required to fulfill this condition. It
follows that
\begin{eqnarray}\label{Delcond}
   \Delta\bigg\langle {\alpha_s\over\pi}\,G^2\bigg\rangle_{\rm ren}
   &=& - {12\Lambda_{\rm V}^4\over\pi^2\beta_0}
    \simeq -0.006~\mbox{GeV}^4 \,, \nonumber\\
   \Delta\Big[ \varepsilon\langle\bar\psi\psi\rangle^2 \Big]_{\rm
ren}
   &=& {\Lambda_{\rm V}^6\over 24\pi^4\beta_0}
    \simeq 5\times 10^{-7}~\mbox{GeV}^6 \,,
\end{eqnarray}
where we have used $\Lambda_{\rm V}=0.461$~GeV corresponding to
$\alpha_s(m_\tau^2)=0.32$. Numerically, the ambiguity of the gluon
condensate is about one third of its value; the ambiguity of the
dimension-six condensate is about $3\times 10^{-3}/\varepsilon$ times
its value, which amounts to 30\% if $\varepsilon=1\%$.

Once the ambiguities cancel for the euclidean correlation function,
they also cancel for the moments. We find that to order $1/m_\tau^6$
the conditions for this cancellation read
\begin{eqnarray}
   \Delta M_k^{\rm ren} + {1\over 3}\,{k+1\over k-2}\,
   {\Delta\big[ \nabla\,\langle O_6^{(1)}\rangle \big]_{\rm ren}
   \over m_\tau^6} &=& 0 \,;\quad k\ne 1,2 \,, \nonumber\\
   \Delta M_1^{\rm ren} - {\Delta\langle O_4^{(1)}\rangle_{\rm ren}
   \over m_\tau^4} - {2\over 3}\,
   {\Delta\big[ \nabla\,\langle O_6^{(1)}\rangle \big]_{\rm ren}
   \over m_\tau^6} &=& 0 \,, \nonumber\\
   \Delta M_2^{\rm ren} + {\Delta\langle O_6^{(1)}(m_\tau^2)
   \rangle_{\rm ren}\over m_\tau^6} + {1\over 3}\,{\Delta\big[
   \nabla\,\langle O_6^{(1)}\rangle \big]_{\rm ren}
   \over m_\tau^6} &=& 0 \,,
\end{eqnarray}
where the perturbative ambiguities $\Delta M_k^{\rm ren}$ have been
given in (\ref{Mkren}). These conditions are indeed satisfied with
(\ref{Delcond}).

\section{Discussion of the results}
\label{sec:7}

We have presented a new analysis of the spectral moments measurable
in hadronic $\tau$ decays, with the main focus on the uncertainties
inherent in the perturbative calculation of the leading terms in the
OPE. Comparing fixed-order perturbation theory (which is known to
order $\alpha_s^3$ in this case) with a partial resummation of the
perturbative series based on the analysis of so-called renormalon
chains, we conclude that the uncertainties in the calculation of the
moments are of order a few per cent. We have also considered the
resummation prescription of Le~Diberder and Pich \cite{LP1} and find
that it does not reduce this uncertainty in a significant way. These
observations imply important limitations for tests of QCD at the
level of power corrections. We have presented a model-independent
description of such corrections and argued that they have to be
included, for reasons of consistency, in order to cure certain
ambiguities of perturbation theory, which are related to IR
renormalons. For the power corrections of dimension four and six we
have demonstrated in detail how these ambiguities can be absorbed
into a redefinition of some nonperturbative parameters (vacuum
condensates). Based on our results, we conclude that previous
analyses of spectral moments in hadronic $\tau$ decays
\cite{LP2,ALEPH,CLEO}, which were aiming at an extraction of power
corrections up to dimension six and even eight, have underestimated
the theoretical uncertainties. In the remainder of this section we
shall reconsider the feasibility of extracting fundamental QCD
parameters, such as the running coupling constant and some of the
vacuum condensates, in the light of our results.

Even in the ideal case in which the spectral moments ${\cal
M}_k^{(J)}$ can be measured with arbitrary precision, and in which
the perturbative contributions to these moments can be calculated to
very high order, there is an unavoidable ambiguity in the definition
of power corrections, which results from the presence of renormalons
in perturbative QCD. The renormalon ambiguities in the values of the
gluon condensate and the dimension-six condensate have been given in
(\ref{Delcond}). They amount to about 30\% of the expected values of
the condensates. However, to achieve such a level of precision would
not only require zero experimental errors, but also a control of the
asymptotic behaviour of perturbation theory, which is lacking at
present. One cannot trust the large-$\beta_0$ limit to provide a very
accurate description of this behaviour. To be conservative, we shall
consider the difference between third-order truncated perturbation
theory and the resummed results in the large-$\beta_0$ limit as an
estimate of the perturbative uncertainty. We will now discuss what
this implies for measurements of $\alpha_s(m_\tau^2)$ and some of the
vacuum condensates.

\begin{figure}[htb]
   \vspace{0.5cm}
   \epsfysize=6cm
   \centerline{\epsffile{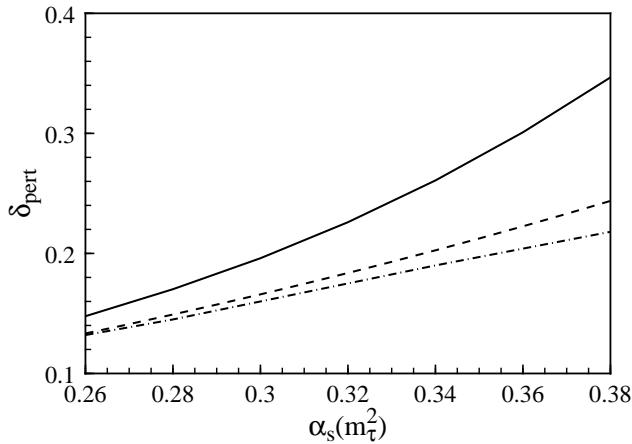}}
   \centerline{\parbox{13cm}{\caption{\label{fig:4}\protect\small\sc
Different perturbative approximations for the quantity $\delta_{\rm
pert}$: resummation in the large-$\beta_0$ limit (solid line), exact
order-$\alpha_s^3$ result in the $\overline{\rm MS}$ scheme (dashed
line), resummation of Le~Diberder and Pich (dash-dotted line).}}}
\end{figure}

The most accurate way to determine the running coupling constant is
to measure the total inclusive ratio $R_\tau$ defined in
(\ref{Rtaudef}). We separate the perturbative contribution to this
quantity from nonperturbative corrections by writing
\begin{equation}
   R_\tau = 3 S_{\rm EW}\,(1 + \delta_{\rm pert}
   + \delta_{\rm power}) \,.
\end{equation}
The perturbative contribution, $\delta_{\rm pert}=2 M_0-2 M_2+M_3-1$,
depends strongly on the value of the running coupling constant. In
Fig.~\ref{fig:4}, we compare the exact order-$\alpha_s^3$ prediction
for this quantity (Table~\ref{tab:1}) to the result obtained from the
resummation of renormalon chains (lower portion of
Table~\ref{tab:4}). For comparison, we also show the result obtained
using the resummation method of Le~Diberder and Pich \cite{LP1}
(Table~\ref{tab:2}), which is routinely used in the determinations of
$\alpha_s$ from $R_\tau$. The differences between these perturbative
approximations are of order few to several per cent, increasing as
the value of $\alpha_s$ increases. The quantity
\begin{eqnarray}
   \delta_{\rm power} &=& \bigg( 2 + {\nabla\over 3} \bigg)\,
    {\langle O_2^{(0+1)}\rangle\over m_\tau^2}
    + 3\,{\nabla\,\langle O_4^{(0+1)}\rangle\over m_\tau^4}
    - 2\,{\langle O_6^{(0+1)}\rangle\over m_\tau^6} \nonumber\\
   &&\mbox{}- \bigg( {4\over 3}\,\nabla - {10\over 9}\,\nabla^2
    \bigg)\,{\langle O_2^{(0)}\rangle\over m_\tau^2}
    + 2 (1-\nabla)\,{\langle O_4^{(0)}\rangle\over m_\tau^4}
    + \dots \simeq - (1.4\pm 0.5)\%
\end{eqnarray}
contains the power corrections. In the SVZ approach, the largest
contribution (about 50\%) comes from dimension-six operators; the
contribution of the gluon condensate is suppressed by two powers of
$\alpha_s$. The remaining corrections come mainly from quark-mass
effects and from the quark condensate, each of which contribute about
25\%. The fact that the power corrections to $R_\tau$ are very small
has been noted in Ref.~\cite{Rtau1}. It is for this reason that
measurements of $R_\tau$ are believed to provide a reliable
determination of the running coupling constant. Note that the
uncertainty in the value of the power corrections is much less than
the uncertainty in the perturbative calculation of $\delta_{\rm
pert}$. Thus, it is the perturbative uncertainty which limits the
accuracy in the extraction of $\alpha_s(m_\tau^2)$. Even if one could
determine $\delta_{\rm pert}$ with a very small error (in practice
this error cannot be much smaller than 1\% because of the theoretical
uncertainty in $\delta_{\rm power}$), there is a rather large
uncertainty in the corresponding value of the running coupling
constant, depending on which perturbative approximation one uses. For
$\delta_{\rm pert}\simeq 0.2$, which is the value preferred by
experiments \cite{ALEPH,CLEO}, one obtains $\alpha_s(m_\tau^2)\simeq
0.303$ from the resummation of renormalon chains,
$\alpha_s(m_\tau^2)\simeq 0.337$ from the exact order-$\alpha_s^3$
calculation, and $\alpha_s(m_\tau^2)\simeq 0.353$ from the
resummation of Le~Diberder and Pich. Hence, we conclude that
\begin{equation}
   \delta\alpha_s(m_\tau^2)\simeq 0.05
\end{equation}
is a reasonable estimate of the theoretical uncertainty. This is a
factor 3 larger than the total error estimated in Ref.~\cite{LP1,LP2}
and quoted in the experimental analyses \cite{ALEPH,CLEO}.

Next consider the extraction of the leading nonperturbative parameter
contributing to the moments, which is the gluon condensate or, more
generally, the quantity $\langle O_4^{(0+1)}\rangle$. From
(\ref{Mkpower}) it follows that the moment ${\cal M}_1^{(0+1)}$ is
most sensitive to this parameter. Let us assume an ideal experiment,
in which this moment can be extracted with high accuracy.\footnote{
In practice, this would require to subtract the contributions with
$J=0$. Moreover, what is accessible in experiments are not the
spectral moments ${\cal M}_k^{(J)}$ themselves, but rather the
moments $R_k$ of the $\tau$ hadronic spectrum, which according to
(\protect\ref{Rkmess}) are combinations of the spectral moments.}
Moreover, assume that $\alpha_s(m_\tau^2)$ has been determined
independently from a measurement of $R_\tau$. As previously, we write
\begin{equation}
   {\cal M}_1^{(0+1)} = 1 + \eta_{\rm pert} + \eta_{\rm power} \,.
\end{equation}
The theoretical predictions for the perturbative contribution,
$\eta_{\rm pert}=M_1-1$, can be obtained directly from the numbers
given in Tables~\ref{tab:1}, \ref{tab:2} and \ref{tab:4}. For values
of $\alpha_s(m_\tau^2)$ in the range from 0.30 to 0.34, the results
differ by 1--3\%. For the leading power corrections we obtain,
neglecting terms of order $10^{-4}$,
\begin{equation}
   \eta_{\rm power} = - {\langle O_4^{(0+1)}\rangle\over m_\tau^4}
   + \dots \simeq 0.2\% - {2\pi^2\over 3 m_\tau^4}\,\bigg( 1
   - {11\over 18}\,{\alpha_s\over\pi} \bigg)\,
   \bigg\langle {\alpha_s\over\pi}\,G^2\bigg\rangle \,.
\end{equation}
For $\langle\frac{\alpha_s}{\pi}\,G^2\rangle \simeq 0.018$~GeV$^2$,
the contribution of the gluon condensate is $-1.1\%$, thus providing
the dominant correction. The precision in the determination of the
gluon condensate is limited by the theoretical uncertainty in the
value of $\eta_{\rm pert}$. An uncertainty of 1\% in the value of
$\eta_{\rm pert}$ implies an uncertainty of 0.02~GeV$^4$ in the value
of the gluon condensate, which is of the same magnitude as the
expected value of the condensate itself. We conclude that, unless one
manages to suppress the perturbative uncertainty, the gluon
condensate cannot be extracted with a precision of much better than
100\%.

Consider finally the extraction of power corrections of dimension
six. An ideal observable to determine these would be the combination
\begin{equation}
   3 {\cal M}_3^{(0+1)} - 2 {\cal M}_2^{(0+1)}
   = 1 + \rho_{\rm pert} + \rho_{\rm power} \,,
\end{equation}
since it does not receive power corrections from $\langle
O_4^{(0+1)}\rangle$ and $\nabla\,\langle O_4^{(0+1)}\rangle$.
Neglecting a contribution of order $10^{-4}$ from $\nabla\,\langle
O_2^{(0+1)}\rangle$, we find
\begin{equation}
   \rho_{\rm power} = - 2\,{\langle O_6^{(0+1)}\rangle\over m_\tau^6}
   + \dots \,.
\end{equation}
In this case, the perturbative uncertainty in the value of $\rho_{\rm
pert}=3 M_3-2 M_2-1$ is of order 1--2\%, leading to $\delta\langle
O_6^{(0+1)}\rangle \simeq 0.2\mbox{--}0.3~\mbox{GeV}^6$, which is
larger than expected value of the condensate. Given the present level
of control over the perturbative contributions to the moments, it is
thus not possible to extract power corrections of dimension six or
higher from hadronic $\tau$ decays, even in an ideal experiment.

To summarize, we believe that beyond extracting the running coupling
constant $\alpha_s(m_\tau^2)$ the goals of a realistic analysis of
spectral moments in hadronic $\tau$ decays should be (i) to test QCD,
or more specifically the SVZ approach, by checking the absence of
large dimension-two operators; (ii) to extract a value of the gluon
condensate. To pursue these goals, we propose to measure differences
of the moments $R_k$ defined in (\ref{Rkmess}). The perturbative
contributions to these differences start at order $\alpha_s^2$ and
are known exactly to order $\alpha_s^4$. Since these contributions
are very small, the quantities we shall construct provide a more
direct measurement of power corrections, which is very little
affected by uncertainties in the value of $\alpha_s(m_\tau^2)$.

A quantity that is sensitive to the presence of non-standard power
corrections of dimension two is
\begin{eqnarray}
   D_2 &=& {1\over 2}\,(R_0 - R_1) = {1\over 6 S_{\rm EW}}
    \int\limits_0^{m_\tau^2}\!{\rm d}s\,\bigg( 1 - {10\over 3}\,
    {s\over m_\tau^2} \bigg)\,{{\rm d}R_\tau(s)\over{\rm d}s}
    \nonumber\\
   &=& {\cal M}_0^{(0+1)} - {5\over 3}\,{\cal M}_1^{(0+1)}
    - {\cal M}_2^{(0+1)} + 3 {\cal M}_3^{(0+1)}
    - {4\over 3}\,{\cal M}_4^{(0+1)} \nonumber\\
   &&\mbox{}- {\cal M}_1^{(0)} + {32\over 9}\,{\cal M}_2^{(0)}
    - {23\over 6}\,{\cal M}_3^{(0)} + {4\over 3}\,{\cal M}_4^{(0)}
    \,.
\end{eqnarray}
Using (\ref{dcoef}), we find that in the $\overline{\rm MS}$ scheme
the perturbative contribution to this difference is
\begin{equation}
   D_2^{\rm pert} = {57\over 80}\,\bigg( {\alpha_s\over\pi} \bigg)^2
   + c_1\,\bigg( {\alpha_s\over\pi} \bigg)^3
   + c_2\,\bigg( {\alpha_s\over\pi} \bigg)^4 + \dots
   \simeq (2.8\pm 1.0) \% \,,
\end{equation}
where
\begin{eqnarray}
   c_1 &=& {241531\over 9600} - {513\over 40}\,\zeta(3)
    \simeq 9.7431 \,, \nonumber\\
   c_2 &=& {596310293\over 768000} - {4617\over 1280}\,\pi^2
    - {1935519\over 3200}\,\zeta(3) + {2565\over 32}\,\zeta(5)
    \simeq 96.898 \,,
\end{eqnarray}
and we have used $\alpha_s=\alpha_s(m_\tau^2)=0.32$. Numerically, we
find that every term in this series is of the same magnitude, and we
have used the last term to estimate the uncertainty. If we perform a
partial resummation of the series in the large-$\beta_0$ limit and
correct for the known pieces to order $\alpha_s^4$ (this is a small
corrections in this case), we find the larger value $D_2^{\rm Borel}
= (5.5\pm 0.7)\%$, where the error is given by the ambiguity due to
the nearest IR renormalon. The leading power corrections of dimension
less or equal to six are
\begin{eqnarray}
   D_2^{\rm power} &=& \bigg( 1 - {3\over 2}\,\nabla \bigg)\,
    {\langle O_2^{(0+1)}\rangle\over m_\tau^2}
    + \bigg( {5\over 3} + {17\over 18}\,\nabla \bigg)\,
    {\langle O_4^{(0+1)}\rangle\over m_\tau^4}
    - {\langle O_6^{(0+1)}\rangle\over m_\tau^6} \nonumber\\
   &&\mbox{}- \bigg( {\nabla\over 9} - {55\over 108}\,\nabla^2
    \bigg)\,{\langle O_2^{(0)}\rangle\over m_\tau^2}
    + \bigg( 1 - {19\over 9}\,\nabla \bigg)\,
    {\langle O_4^{(0)}\rangle\over m_\tau^4} + \dots \,.
\end{eqnarray}
In the SVZ approach the largest contribution comes from the
gluon condensate, and we find $D_2^{\rm power}\simeq (0.8\pm 1.0)\%$.
We conclude that
\begin{equation}
   D_2\simeq\cases{ (3.6\pm 1.3)\% \,;
    &fixed-order perturbation theory, \cr
   (6.3\pm 1.2)\% \,; &resummed perturbation theory. \cr}
\end{equation}
A non-standard dimension-two contribution of size
$(\Lambda/m_\tau)^2\sim 3$--8\% (for $\Lambda=300$--500~MeV) and
possibly negative sign could be large enough to change this value in
a significant way. Such a contribution is forbidden in the standard
SVZ approach, since there is no local, gauge-invariant operator of
dimension two in QCD. We note, however, that the truncation of the
perturbative series for $D_2$ can fake such a term, because this
series contains an UV renormalon at $u=-1$ \cite{Alta}. It would be
interesting to see whether the resummation of renormalon chains,
which resums this contribution in the large-$\beta_0$ limit, improves
the agreement with experiment.

Another interesting quantity, which can be used to measure the gluon
condensate, is
\begin{eqnarray}
   D_4 &=& {3\over 10}\,(R_2 - R_1) = {1\over 3 S_{\rm EW}}
    \int\limits_0^{m_\tau^2}\!{\rm d}s\,{s\over m_\tau^2}\,\bigg(
    {9\over 4}\,{s\over m_\tau^2} - 1 \bigg)\,
    {{\rm d}R_\tau(s)\over{\rm d}s} \nonumber\\
   &=& - {\cal M}_1^{(0+1)} + {3\over 2}\,{\cal M}_2^{(0+1)}
    + {3\over 2}\,{\cal M}_3^{(0+1)} - {7\over 2}\,{\cal M}_4^{(0+1)}
    + {3\over 2}\,{\cal M}_5^{(0+1)} \nonumber\\
   &&\mbox{}+ {4\over 3}\,{\cal M}_2^{(0)} - {17\over 4}\,
    {\cal M}_3^{(0)} + {22\over 5}\,{\cal M}_4^{(0)}
    - {3\over 2}\,{\cal M}_5^{(0)} \,.
\end{eqnarray}
The perturbative contribution to this difference is
\begin{equation}
   D_4^{\rm pert} = - {27\over 160}\,\bigg( {\alpha_s\over\pi}
   \bigg)^2 - d_1\,\bigg( {\alpha_s\over\pi} \bigg)^3
   - d_2\,\bigg( {\alpha_s\over\pi} \bigg)^4 + \dots
   \simeq - (0.46\pm 0.10) \% \,,
\end{equation}
with
\begin{eqnarray}
   d_1 &=& {34527\over 6400} - {243\over 80}\,\zeta(3)
    \simeq 1.7436 \,, \nonumber\\
   d_2 &=& {77871081\over 512000} - {2187\over 2560}\,\pi^2
    - {819369\over 6400}\,\zeta(3) + {1215\over 64}\,\zeta(5)
    \simeq 9.4508 \,.
\end{eqnarray}
The resummation of renormalon chains gives a very similar result,
$D_4^{\rm Borel}=-(0.55\pm 0.40)\%$. In both cases the perturbative
contribution is very small, and the accuracy seems to be limited by
the renormalon ambiguity. We also note that the coefficient of the UV
renormalon at $u=-1$ is very small in this case, so that one does not
expect large uncertainties due to the truncation of the series. The
leading power corrections to the quantity $D_4$ are
\begin{eqnarray}
   D_4^{\rm power} &=& - {13\over 40}\,
    {\nabla\,\langle O_2^{(0+1)}\rangle\over m_\tau^2}
    + \bigg( 1 - {35\over 24}\,\nabla \bigg)\,
    {\langle O_4^{(0+1)}\rangle\over m_\tau^4}
    + {3\over 2}\,{\langle O_6^{(0+1)}\rangle\over m_\tau^6}
    \nonumber\\
   &&\mbox{}+ \bigg( {\nabla\over 30} - {167\over 1800}\,\nabla^2
    \bigg)\,{\langle O_2^{(0)}\rangle\over m_\tau^2}
    - {7\over 24}\,{\nabla\,\langle O_4^{(0)}\rangle\over m_\tau^4}
    + \dots \nonumber\\
   &\simeq& 0.3\% + {2\pi^2\over 3 m_\tau^4}\,\bigg( 1
    - {11\over 18}\,{\alpha_s\over\pi} \bigg)\,
    \bigg\langle {\alpha_s\over\pi}\,G^2\bigg\rangle \,.
\end{eqnarray}
To good approximation, they are given by the gluon condensate.
Combining them with the perturbative contribution, we obtain
\begin{equation}
   D_4\simeq 1.1\%\times {\bigg\langle\displaystyle
   {\alpha_s\over\pi}\,G^2\bigg\rangle\over 0.018~\mbox{GeV}^4}
   - (0.2\pm 0.4)\% \,.
\end{equation}
Hence, a measurement of $D_4$ comes close to a ``null measurement''.
If a value $D_4\ne 0$ can be established, it provides direct evidence
for the existence of the gluon condensate. We would consider such a
measurement more convincing than the existing fits \cite{ALEPH,CLEO}
of the gluon condensate from a moment analysis following the lines of
Ref.~\cite{LP2}.

\newpage
\section*{Appendices}
\appendix

\section{Calculation of the integrals $I_n^{k+1}(m_\tau)$}
\label{app:int}

Le Diberder and Pich have introduced the integrals \cite{LP2}
\begin{eqnarray}
   I_n^{k+1}(m_\tau) &=& {1\over 2\pi i}\,\oint\limits_{|x|=1}
    {{\rm d}x\over x}\,(1 - x^{k+1})\,\bigg(
    {\alpha_s(-x m_\tau^2)\over\pi} \bigg)^n \nonumber\\
   &=& {1\over 2\pi} \int\limits_{-\pi}^\pi\!{\rm d}\varphi\,
    \Big[ 1 + (-1)^k\,e^{i(k+1)\varphi} \Big]\,\bigg(
    {\alpha_s(e^{i\varphi} m_\tau^2)\over\pi} \bigg)^n \,.
\end{eqnarray}
The running coupling constant along the contour in the complex plane
can be obtained from the solution of the RGE (\ref{RGE}), which reads
\begin{equation}
   i\varphi\,{\beta_0\over 4\pi} = {1\over\alpha_s(\varphi)}
   - {1\over\alpha_s(0)} + {\beta_1\over 4\pi\beta_0}\,
   \ln{\alpha_s(\varphi)\over\alpha_s(0)}
   + {\beta_0\beta_2-\beta_1^2\over 16\pi^2\beta_0^2}\,
   \Big[ \alpha_s(\varphi) - \alpha_s(0) \Big] + \dots \,,
\end{equation}
with $\alpha_s(\varphi)\equiv\alpha_s(e^{i\varphi} m_\tau^2)$.

In the approximation where one uses the one-loop $\beta$-function
(setting $\beta_1=\beta_2=\dots=0$), it is possible to perform the
contour integrals explicitly.
The result for $n=1$ is
\begin{equation}
   I_1^{k+1}(m_\tau) = {4\over\beta_0}\,\bigg\{ {(-1)^k\over\pi}\,
   e^{-(k+1)/a}\,\mbox{Im}\,\mbox{Ei}\Big[ (k+1)(1/a + i\pi)
   \Big] + {1\over\pi}\,\arctan(\pi a) \bigg\} \,,
\end{equation}
where
\begin{equation}
   a = {\beta_0\over 4\pi}\,\alpha_s(m_\tau^2)
   = {1\over\ln(m_\tau^2/\Lambda^2)} \,,
\end{equation}
and $\mbox{Ei}(x)=\int_{-\infty}^x\frac{{\rm d}t}{t}\,e^t$ is the
exponential integral. Note that the value of
$\mbox{Im}\,\mbox{Ei}(x+iy)$ is not affected by the pole at $t=0$.
The integrals with $n>1$ can be obtained from the recursion relation
\begin{equation}
   I_n^{k+1}(m_\tau) = \bigg( {4\over\beta_0} \bigg)^{n-1}\,
   {1\over(n-1)!}\,\bigg(-{{\rm d}\over{\rm d}a^{-1}} \bigg)^{n-1}\,
   I_1^{k+1}(m_\tau) \,.
\end{equation}

Beyond this approximation, the integrals have to be performed
numerically. However, in the case of the two-loop $\beta$-function
one can still obtain an approximate analytic expression by writing
\begin{eqnarray}
   i\varphi\,{\beta_0\over 4\pi} &=& {1\over\alpha_s(\varphi)}
    - {1\over\alpha_s(0)} + {\beta_1\over 4\pi\beta_0}\,
    \ln{\alpha_s(\varphi)\over\alpha_s(0)} \nonumber\\
   &\simeq& {1\over\alpha_s(\varphi)} - {1\over\alpha_s(0)}
    - i\varphi\,{\beta_1\over 16\pi^2}\,\alpha_s(0) \,.
\end{eqnarray}
This leads to the simple relation
\begin{equation}\label{Inkappr}
   I_n^{k+1}\Big|_{\rm 2-loop} \simeq \bigg( 1
   + {\beta_1\over\beta_0^2}\,a \bigg)^{-n}\,
   I_n^{k+1}\Big|_{\rm 1-loop}(a^*) \,,
\end{equation}
where in the one-loop integrals one uses
\begin{equation}
   a^* = \bigg( 1 + {\beta_1\over\beta_0^2}\,a \bigg)\,a
\end{equation}
instead of $a$. This provides a very good approximation to the exact
numerical results.

In Table~\ref{tab:5}, we compare the various approximations for some
of these integrals in the case of $\alpha_s(m_\tau^2)=0.32$ (in the
$\overline{\rm MS}$ scheme).

\begin{table}[t]
\centerline{\parbox{15cm}{\caption{\label{tab:5}\protect\small\sc
Integrals $I_n^{k+1}$ evaluated using the one-loop (upper portion),
two-loop (middle portion), and three-loop (lower portion)
$\beta$-function. In the two-loop case, the approximate results
obtained from (\protect\ref{Inkappr}) are given below the exact
values.}}}
\vspace{0.5cm}
\centerline{\begin{tabular}{c|rrrrrr}
\hline\hline
\rule[-0.2cm]{0cm}{0.7cm} & $k=0$ & $k=1$ & $k=2$ & $k=3$ & $k=4$ &
 $k=5$ \\
\hline
\rule[-0.1cm]{0cm}{0.5cm}
$I_1^{k+1}~(\times 10^{-2})$ & 11.55 & 9.73 & 9.40 & 9.24 & 9.16 &
 9.10 \\
\rule[-0.1cm]{0cm}{0.5cm}
$I_2^{k+1}~(\times 10^{-3})$ & 12.08 &  8.01 & 7.62 & 7.40 & 7.28 &
 7.20 \\
\rule[-0.1cm]{0cm}{0.5cm}
$I_3^{k+1}~(\times 10^{-4})$ & 11.67 &  5.25 & 5.22 & 5.05 & 4.95 &
 4.89 \\
\hline
\rule[-0.1cm]{0cm}{0.5cm}
$I_1^{k+1}~(\times 10^{-2})$ & 11.44 & 9.17 & 8.91 & 8.75 & 8.67 &
 8.61 \\
approx. & 11.57 & 9.37 & 9.04 & 8.88 & 8.78 & 8.72 \\
\rule[-0.1cm]{0cm}{0.5cm}
$I_2^{k+1}~(\times 10^{-3})$ & 11.54 & 6.68 & 6.65 & 6.44 & 6.34 &
 6.28 \\
approx. & 11.79 & 6.95 & 6.74 & 6.54 & 6.43 & 6.36 \\
\rule[-0.1cm]{0cm}{0.5cm}
$I_3^{k+1}~(\times 10^{-4})$ & 10.56 & 3.34 & 4.12 & 3.92 & 3.86 &
 3.82 \\
approx. & 10.86 & 3.51 & 4.03 & 3.89 & 3.82 & 3.78 \\
\hline
\rule[-0.1cm]{0cm}{0.5cm}
$I_1^{k+1}~(\times 10^{-2})$ & 11.42 & 9.11 & 8.87 & 8.71 & 8.63 &
 8.57 \\
\rule[-0.1cm]{0cm}{0.5cm}
$I_2^{k+1}~(\times 10^{-3})$ & 11.48 & 6.58 & 6.59 & 6.38 & 6.28 &
 6.22 \\
\rule[-0.1cm]{0cm}{0.5cm}
$I_3^{k+1}~(\times 10^{-4})$ & 10.46 & 3.21 & 4.08 & 3.85 & 3.80 &
 3.76 \\
\hline\hline
\end{tabular}}
\vspace{0.5cm}
\end{table}

\section{Asymptotic behaviour of the functions $W_k(\tau)$}
\label{app:asy}

In order to perform numerical integrations with the distribution
functions $W_k(\tau)$ defined in (\ref{W0fun})--(\ref{Wkfuns}), it
is convenient to use their asymptotic behaviour for large and small
values of $\tau$. For $\tau\gg 1$, we find
\begin{eqnarray}
   W_k(\tau) &=& {32\over 3}\,(k+1)\,\Bigg\{ \bigg(
    {5 k+16\over 36(k+2)^2} + {\ln\tau\over 6(k+2)} \bigg)\,
    {1\over\tau} \nonumber\\
   &&\phantom{ {32\over 3}\,(k+1)\,\Bigg\{ }
    - \bigg( {7 k+33\over 144(k+3)^2}
    + {\ln\tau\over 12(k+3)} \bigg)\,{1\over\tau^2}
    + O(\tau^{-3}) \Bigg\} \,,
\end{eqnarray}
which is valid for all $k\ge 0$. For $\tau\ll 1$, on the other hand,
we obtain
\begin{eqnarray}
   W_0(\tau) &=& {32\over 3}\,\Bigg\{ \Big( 4 - 3\zeta(3) \Big)\,\tau
    - {3\over 4}\,\tau^2 + \bigg( {1\over 2}
    - {1\over 4}\,\ln\tau \bigg)\,\tau^3 + O(\tau^4) \Bigg\} \,,
    \nonumber\\
   W_1(\tau) &=& {32\over 3}\,\Bigg\{ \bigg( 5 - 6\zeta(3)
    - {3\over 2}\,\ln\tau \bigg)\,\tau^2 + \bigg( {5\over 2}
    - \ln\tau \bigg)\,\tau^3 + O(\tau^4) \Bigg\} \,, \nonumber\\
   W_2(\tau) &=& {32\over 3}\,\Bigg\{ {9\over 4}\,\tau^2
    - \bigg( {4\over 3} - {9\over 4}\,\ln\tau + {3\over 4}\,
    \ln^2\!\tau \bigg)\,\tau^3 + O(\tau^4) \Bigg\} \,,
\end{eqnarray}
as well as
\begin{equation}
   W_k(\tau) = {32\over 3}\,(k+1)\,\Bigg\{ {3\over 4(k-1)}\,\tau^2
    - \bigg( {3 k-8\over 4(k-2)^2} - {\ln\tau\over 2(k-2)}
    \bigg)\,\tau^3 + O(\tau^4) \Bigg\}
\end{equation}
for $k\ge 3$.

The asymptotic behaviour of the distribution function $w_D(\tau)$ in
(\ref{wDfun}) can be derived from the above results by taking the
limit $k\to\infty$. For $\tau\gg 1$, we find
\begin{equation}
   w_D(\tau) = {16\over 9}\,\bigg( \ln\tau + {5\over 6} \bigg)\,
   {1\over\tau} - {8\over 9}\,\bigg( \ln\tau + {7\over 12} \bigg)\,
   {1\over\tau^2} + O(\tau^{-3}) \,,
\end{equation}
whereas for $\tau\ll 1$
\begin{equation}
   w_D(\tau) = 8\tau^2 + 8\,\bigg( {2\over 3}\,\ln\tau
   - 1 \bigg)\,\tau^3 + O(\tau^4) \,.
\end{equation}

\section{QCD parameters}
\label{app:QCD}

For the running quark masses at the scale $\mu_0=1$~GeV (in the
$\overline{\rm MS}$ scheme), we use
\begin{eqnarray}
   m_u(\mu_0^2) &=& 5\pm 1~\mbox{MeV} \,, \nonumber\\
   m_d(\mu_0^2) &=& 9\pm 1~\mbox{MeV} \,, \nonumber\\
   m_s(\mu_0^2) &=& 178\pm 18~\mbox{MeV} \,.
\end{eqnarray}
The value for $m_s$ has been obtained in a recent QCD sum rule
analysis \cite{JaMue}. The scale dependence of the running masses is
governed by the RGE
\begin{equation}
   {{\rm d} m_q^2(\mu^2)\over{\rm d}\ln\mu^2}
   = \gamma_m[\alpha_s(\mu^2)]\,m_q^2(\mu^2) \,.
\end{equation}
At two-loop order,
\begin{equation}
   \gamma_m(\alpha_s) = \gamma_0\,{\alpha_s\over 4\pi}
    + \gamma_1\,\bigg( {\alpha_s\over 4\pi} \bigg)^2 + \dots \,,
\end{equation}
with coefficients $\gamma_0=-8$ and $\gamma_1=-364/3$ (for $n_f=3$)
\cite{Tarr}.
At this order, the solution of the RGE is
\begin{equation}
   m_q^2(\mu^2) = m_q^2(\mu_0^2)\,\Bigg(
   {\alpha_s(\mu_0^2)\over\alpha_s(\mu^2)}
   \Bigg)^{\gamma_0/\beta_0}\,\bigg\{ 1
   + {\beta_0\gamma_1 - \beta_1\gamma_0\over 4\beta_0^2}\,
   {\alpha_s(\mu_0^2) - \alpha_s(\mu^2)\over\pi} + \dots \bigg\} \,.
\end{equation}
This can be used to calculate $m_q(m_\tau^2)$. In the numerical
analysis we use $\alpha_s(m_\tau^2)=0.32$, unless otherwise
specified.

In the calculation of the power corrections in Sect.~\ref{sec:5} we
also need the elements $V_{uj}$ of the Cabibbo-Kobayashi-Maskawa
matrix. We neglect $V_{ub}$, so that $|\,V_{ud}|^2 + |\,V_{us}|^2=1$,
and use
\begin{equation}
   |\,V_{ud}| = 0.9753 \,,\qquad |\,V_{us}| = 0.2209 \,.
\end{equation}

Finally, we give the definition of the scale-invariant condensates of
dimension four, which are relevant to our analysis. In the
$\overline{\rm MS}$ scheme, and to next-to-leading order in the
coupling constant, they are defined as \cite{Gene,Spir}
\begin{eqnarray}
   \bigg\langle {\alpha_s\over\pi}\,G^2 \bigg\rangle
   &=& \bigg( 1 + {16\over 9}\,{\alpha_s(\mu^2)\over\pi} \bigg)\,
    {\alpha_s(\mu^2)\over\pi}\,\langle G_{\mu\nu} G^{\mu\nu}(\mu^2)
    \rangle \nonumber\\
   &&\mbox{}- {16\over 9}\,{\alpha_s(\mu^2)\over\pi}\,\bigg( 1
    + {19\over 24}\,{\alpha_s(\mu^2)\over\pi} \bigg)\,\sum_k\,
    m_k(\mu^2)\,\langle\bar\psi_k\psi_k(\mu^2)
    \rangle \nonumber\\
   &&\mbox{}- {1\over 3\pi^2}\,\bigg( 1 + {4\over 3}\,
    {\alpha_s(\mu^2)\over\pi} \bigg)\,\sum_k\,m_k^4(\mu^2) \,,
    \nonumber\\
   \langle m_i\bar\psi_i\psi_i\rangle &=& m_i(\mu^2)\,
    \langle \bar\psi_i\psi_i(\mu^2) \rangle
    + {3\over 7\pi^2}\,\bigg( {\pi\over\alpha_s(\mu^2)}
    - {53\over 24} \bigg)\,m_i^4(\mu^2) \,.
\end{eqnarray}
For the scale-invariant quark condensates we write $\langle
m_i\bar\psi_i\psi_i\rangle = -m_i(\mu_0^2)\,\kappa_i$ and take the
standard values \cite{SVZ}
\begin{equation}
   \kappa_u = \kappa_d = (230\pm 30~\mbox{MeV})^3 \,,\qquad
   \kappa_s = 0.65\kappa_u \,.
\end{equation}
For the scale-invariant gluon condensate and for the four-quark
condensate we use an average over some recent determinations
\cite{Laun}--\cite{GiBo}, however with conservative errors:
\begin{eqnarray}
   \bigg\langle {\alpha_s\over\pi}\,G^2 \bigg\rangle
   &=& (0.018\pm 0.009)~\mbox{GeV}^4 \,, \nonumber\\
   \phantom{ \bigg[ }
   \rho\alpha_s\langle\bar\psi\psi\rangle^2
   &=& (3.5\pm 2.0)~\mbox{GeV}^6 \,.
\end{eqnarray}

\end{document}